\documentclass[%
 reprint,
 amsmath,amssymb,
 aps,
]{revtex4-2}

\usepackage{graphicx}
\usepackage{dcolumn}
\usepackage{bm}
\usepackage[T1]{fontenc}
\usepackage[utf8]{inputenc}

\DeclareUnicodeCharacter{03B4}{$\delta$}
\DeclareUnicodeCharacter{03B1}{$\alpha$}
\DeclareUnicodeCharacter{03B2}{$\beta$}
\DeclareUnicodeCharacter{03B3}{$\gamma$}

\begin{document}

\preprint{APS/123-QED}

\title{Size dependent solid-solid crystallization of halide perovskites}

\author{Paramvir Ahlawat}
\email{paramvir.chem@gmail.com}
\altaffiliation{Yusuf Hamied Department of Chemistry, University of Cambridge} 

\date{March 25, 2024}

\begin{abstract}
The efficiency and stability of halide perovskite-based solar cells and light-emitting diodes directly depend on the intricate dynamics of solid-solid crystallization\cite{ eperon_inorganic_2015,  mcmeekin_crystallization_2017, liu_high-performance_2017, gratia_many_2017, kim_methylammonium_2019, wang_thermodynamically_2019, lu_vapor-assisted_2020, lee_solid-phase_2020, yi_intermediate-phase-assisted_2020, ahlawat_combined_2021, yoon_surface_2021, park_controlled_2023, tong_scalable_2021,  wang_formation_2021, alharbi_formation_2021,  arora_kinetics_2022, duan_phase-pure_2022, bu_modulating_2022, zhao_inactive_2022, myung_challenges_2022, liu_retarding_2023, mcmeekin_intermediate-phase_2023, ahlawat_crystallization_2023}. In this study, we employ a multi-scale approach using random phase approximation, density functional theory, machine learning potentials, reduced charge force fields, and both enhanced sampling biased and brute-force unbiased molecular dynamics simulations to understand the solid-solid phase transitions in cesium lead iodide perovskite. Our simulations uncover that the direct phase transition from the non-perovskite to the perovskite involves the formation of stacked-faulted and low-dimensional intermediate structures. Through extensive large-scale all-atom simulations encompassing up to 650,000 atoms, we observe that solid-solid crystallization may require the formation of a sufficiently large critical nucleus to grow into a faceted perovskite crystal. Based on simulations, we determine that utilizing (100)-faceted seeded crystallization could offer a promising path for manufacturing high-performance and stable perovskite solar cells.

\end{abstract}

\maketitle

\section{\label{sec:level1}Introduction\protect\\ }

Halide perovskites\cite{kagan_organic-inorganic_1999, kojima_organometal_2009, im_6.5_2011, lee_efficient_2012, kim_lead_2012} are photo-active materials with tunable band gap ($1.1 - 2.3$ eV) to manufacture record efficiency solar cells (PSCs)\cite{bremner_optimum_2016, jeong_pseudo-halide_2021, yoon_surface_2021, chu_surface_2023, noauthor_interactive_nodate, chen_regulating_2023, park_controlled_2023}. Despite their superior optoelectronic properties for making low-priced efficient photovoltaics and light emitting diodes, their wide-spread industrialization is still suffering because of the poor reproducibility and stability using current manufacturing processes. 

\begin{figure}
  \includegraphics[width=0.94\linewidth]{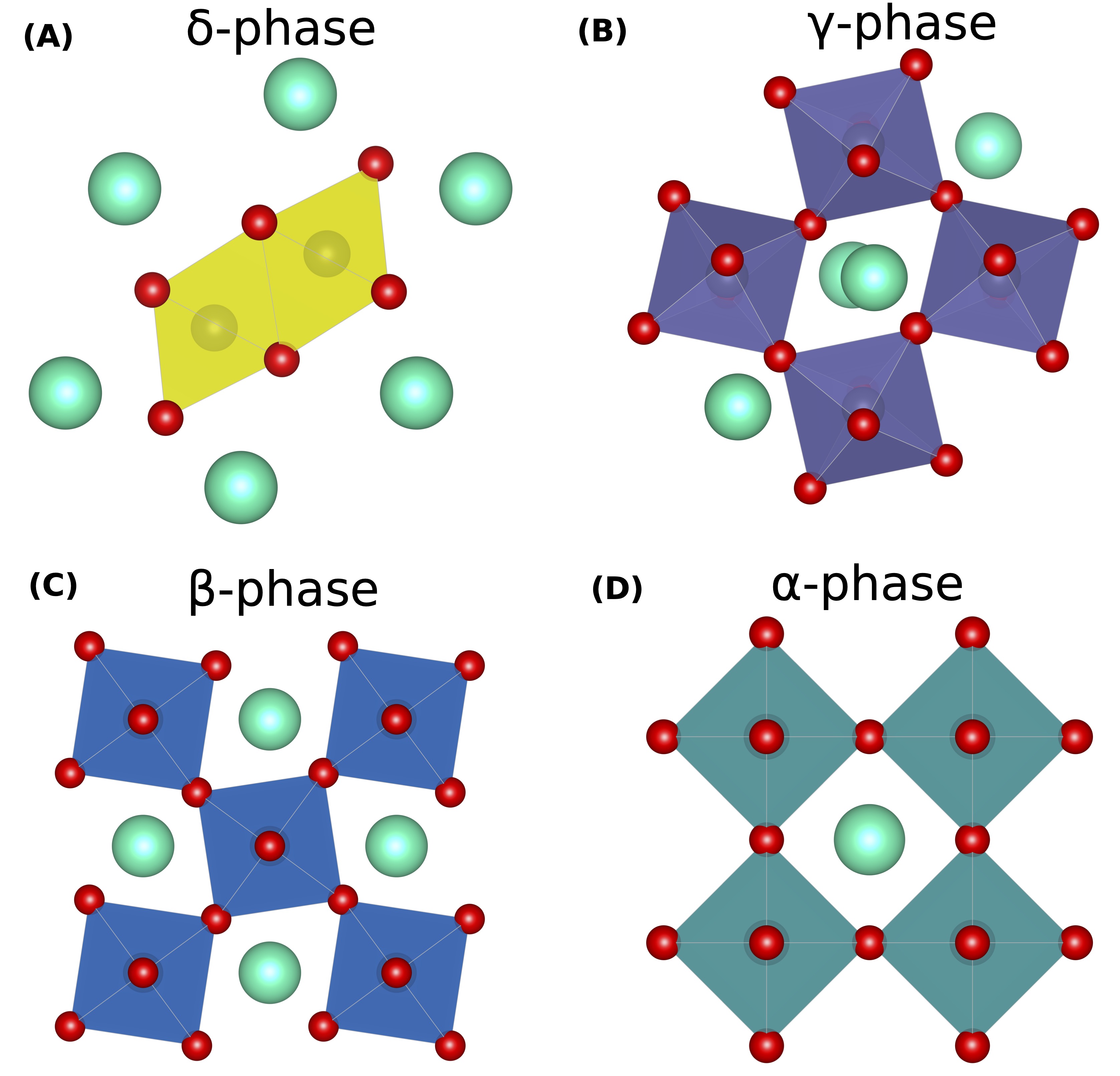} 
  \caption{\label{fig:all} \textbf{Experimentally known polymoprhs of CsPbI\textsubscript{3}}: Figure shows the different phases of CsPbI\textsubscript{3} with Pb-I structures as octahedra with red color iodides on corners and Cs atoms as light-green spheres.}
\end{figure}

Where \textit{\textbf{one of the key problem is the lack of suitable and reproducible synthesis recipes to produce defect free highly stable thin-films of halide perovskites.}} Therefore, it is of key importance to comprehend the details of their formation process which can assist to design better experimental methodologies. In particular, wide range of experiments have established that the solid-solid phase transitions dominate the crystallization process and control the micro-structure evolution of perovskite thin-films. In this work, we perform molecular dynamics simulations and try to understand the details of crystallization of the widely studied and thermally stable cesium lead iodide (CsPbI\textsubscript{3}). \\

CsPbI\textsubscript{3} has four experimentally known polymorphs\cite{moller_crystal_1958, stoumpos_semiconducting_2013, trots_high-temperature_2008, sutton_cubic_2018, eperon_inorganic_2015, jena_stabilization_2018}, often recognized as $\delta$, $\gamma$, $\beta$ and $\alpha$-phases visualized in Figure \ref{fig:all}. Among them, $\delta$-phase is the most stable form, and a yellow-colored photo-inactive material consists of edge-sharing lead-iodide octahedra surrounded by cesium cations. On the other hand, $\gamma$, $\beta$ and $\alpha$-phases are black coloured thermally stable photo-active perovskites made of corner-sharing lead-iodide octahedra with different tilted angles, however meta-stable compared to its $\delta$-form. The perovskite phases have suitable band-gap of $\sim1.7$ to $\sim1.9eV$ for making tandem solar cells with the current leading silicon technology (efficiency $\sim$ 26\%) and theoretically can reach efficiencies beyond 30\%\cite{bremner_optimum_2016}. However, to construct a competitive low priced and stable $>30\%$ efficient and stable perovskite-silicon or all perovskites tandem solar cells \cite{eperon_perovskite-perovskite_2016, chin_interface_2023, jost_textured_2018, mariotti_interface_2023, liu_efficient_2022, noauthor_interactive_nodate, wang_suppressed_2023, ji_stable_2022, chu_surface_2023, mali_phase-heterojunction_2023} involving CsPbI\textsubscript{3}: \textit{\textbf{one of the main quests is to synthesize/crystallize defects-free corner-sharing phases of CsPbI\textsubscript{3} and eliminate/prevent the formation of edge-sharing $\delta$-phase}}. \\

\newpage

\onecolumngrid

\begin{figure}
  \includegraphics[width=0.8\linewidth]{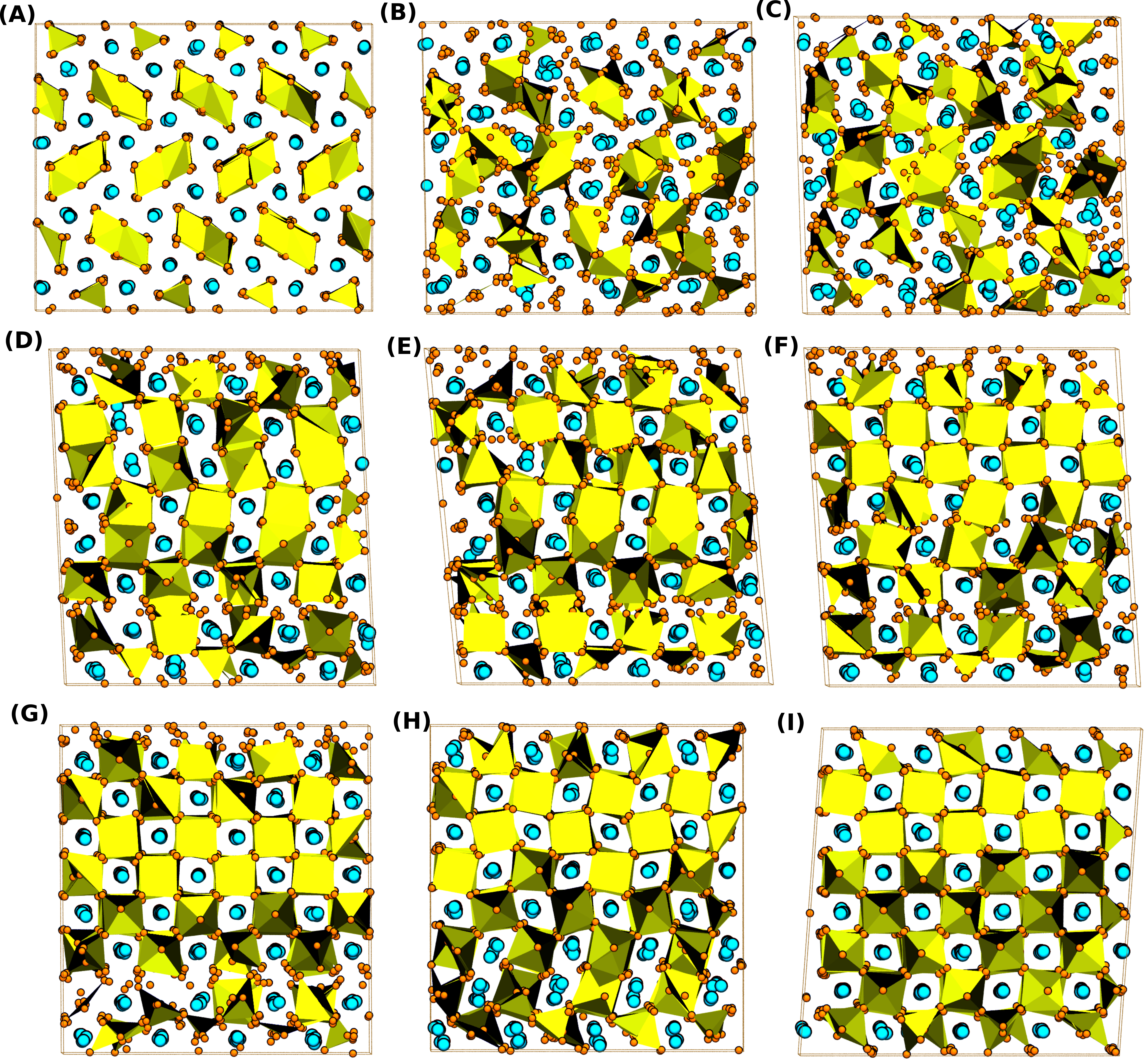}
  \caption{\label{fig:full} \textbf{Solid-solid phase transition by enhanced sampling}: Figure shows the simulated transformation of edge-sharing $\delta$-phase to corner-sharing $\alpha$-phase of CsPbI\textsubscript{3}. Pb-I structures are represented as yellow octahedra with orange color iodides on coners, and Cs atoms are depicted with blue spheres.}
\end{figure}

\twocolumngrid

Absolutely gigantic and strenuous experimental efforts\cite{wang_formation_2021} have been employed to reach the present day synthesis recipes. A wide range of processes and uncountable additives have been tried to improve the crystallization process. \\

Now, it is common to utilize additives like dimethylammonium\cite{ke_myths_2018, meng_chemical_2020, lau_fabrication_2019, wang_thermodynamically_2019}, phenylethylammonium\cite{jiang_reduced-dimensional_2018}, bromide\cite{beal_cesium_2016, li_phase_2017, zhou_light-independent_2017, xiang_europium-doped_2019} and chloride\cite{wang_chlorine_2019, yoon_surface_2021} to make efficient and optimistically stable black-phase of cesium lead iodide. Nevertheless, existing manufacturing processes continue to struggle with issues of reproducibility and stability, prompting ongoing efforts to enhance them through extensive experiments aimed at achieving progressively higher efficiencies and industrial scale stable PSCs. Therefore, it is of key importance to understand and optimize the synthesis routes to ensure the stability and reliability of perovskite-based devices in practical applications. A comprehensive atomic-level picture of the nucleation and growth processes can help understand the current experimental recipes' limitations and, in turn, assist to design improved, reproducible and even newer methodologies for making PSCs and LEDs. \\

Towards this goal, we perform molecular dynamics (MD) simulations of the crystallization of cesium lead iodide. A simulation study of the full-scale experiment is challenging and herculean task. Here, we focus on understanding the solid-solid phase transitions in perovskites which are experimentally known to controls the synthesis process of perovskite thin-films. We explore two key aspects that have not been extensively investigated in the context of phase transitions in halide perovskites. \\

\newpage

At first, we study how the corner-sharing photo-active perovskite black phase originates from the edge-sharing yellow phase. Secondly, we aim to unravel the large scale crystallization mechanism of solid-solid transitions in perovskites. To address these fundamental questions, we employ a unique combination of simulation methods, including random phase approximation (RPA), density functional theory (DFT), machine learning potentials (MLPs), umbrella sampling, and large-scale seeded simulations utilizing reduced charge inter-atomic potentials (RCPs). Through biased simulations using RCPs and MLPs for the direct phase transition from the non-perovskite to perovskite phase, we identify activated pathways characterized by the formation of intermediary stack-faulted structures and edge-dislocations. In our exploration of larger size simulations, we uncover that the solid-solid transitions could adheres to nucleation process. This implies that the formation of a critical nucleus may become essential for facilitating the growth into stable and faceted crystals. We sheds light on the intricate details of the nucleation from one phase to another, providing valuable insights into the underlying fundamental mechanisms. \\


\section{\label{sec:level1}Direct non-perovskite ($\delta$) phase to perovskite phase transition\protect\\ }
In conventional experimental procedures, the thermodynamically stable $\delta$-phase initially crystallizes and subsequently undergoes transformation into the black-phase through annealing at elevated temperatures. However, direct conversion processes have historically yielded less stable and less efficient perovskite solar cells.  Therefore, first we start with the MD simulations of the direct conversion of edge-sharing $\delta$-phase of CsPbI\textsubscript{3} to the corner-sharing photo-active phase. Previously Grünwald, Limmer and co-workers\cite{lin_thermochromic_2018, bischak_liquid-like_2020} have conducted notable co-existing simulations, revealing a liquid-like interface mediating the phase transition between edge-corner-sharing polymorphs of CsPbI\textsubscript{3}. Another significant contribution by Pramchu \textit{et al.}\cite{pramchu_effects_2019} employed variable-cell nudged-elastic band calculations to investigate reverse phase transitions from the corner-sharing to the edge-sharing phase of CsPbI\textsubscript{3}, primarily aiming to gain insights into the degradation process of smaller systems using DFT calculations. \\

However, to understand the crystallization of perovskite black phases, it is necessary to reveal formation pathway of meta-stable perovskite phases from a direct heating the $\delta$-phase as per experimental procedure. Here, we perform MD simulations of the direct phase transition, initiated solely from the $\delta$-CsPbI\textsubscript{3}. First-order phase transitions between crystalline phases are often associated with substantial energy barriers\cite{noauthor_statistical_1980, binder_theory_1987, frenkel_understanding_2002, auer_prediction_2001, du_shape-driven_2017, peters_reaction_2017, xiao_mechanism_2013, qian_variable_2013, frenkel_entropy-driven_1999, qi_nonclassical_2015, grunwald_mechanisms_2006, grunwald_nucleation_2009, zahn_nucleation_2013, khaliullin_nucleation_2011, anwar_polymorphic_2017, du_shape-driven_2017, rogal_neural-network-based_2019, caspersen_finding_2005}, potentially occurring on time scales ranging from seconds to minutes\cite{peng_two-step_2015, sanz_mediated_2015, zhu_revisited_2020, li_revealing_2021, frenkel_structure_1986}. Simulating these experimental crystallization processes within feasible times and resources poses computational challenges. \\

To address this, we employ biased enhanced sampling methods\cite{torrie_nonphysical_1977, frenkel_understanding_2002, virnau_calculation_2004, huber_local_1994, invernizzi_rethinking_2020, debnath_gaussian_2020, maragakis_gaussian-mixture_2009}, specifically umbrella sampling\cite{torrie_nonphysical_1977,  ten_wolde_simulation_1996, auer_prediction_2001} and OPES\cite{invernizzi_rethinking_2020, invernizzi_unified_2020}. These methods explore the free energy surface by applying a bias potential to the potential energy surface. The bias potential is crafted as a function of collective coordinates, commonly referred to as collective variables (CVs). The choice of suitable CVs is crucial\cite{rohrdanz_determination_2011, peters_reaction_2017, mccarty_variational_2017, sultan_automated_2018, rizzi_blind_2019} for addressing complex geometries in multi-species systems, such as the process under investigation. In this study, we introduce a locally ordered CV\cite{lechner_accurate_2008, ahlawat_atomistic_2020} based on the local crystallographic information of perovskite ($s^{\alpha}(r_i^{Pb})$)\cite{niu_molecular_2018, bonati_silicon_2018, ahlawat_combined_2021, ahlawat_molecular_2023} for species ${\alpha}$ = (Cs, Pb, and I), centered around positions \textit{$r_i^{Pb}$} of Pb atoms. \\

\begin{equation}\label{eqn:CV1}
 s^{\alpha}(r_i^{Pb}) = \frac{ 1 - \left(\frac{s(q)_i^{\alpha}}{s(q)^{\alpha}_{cut}}\right)^{-c}} 
 {1 - \left(\frac{s(q)_i^{\alpha}}{s(q)^{\alpha}_{cut}}\right)^{-d}}
\end{equation}
Where $s(q)_i^{\alpha}$ is defined as:
\begin{equation}\label{eqn:CV2}
s(q)_i^{\alpha} = 1 + \frac{1}{N}\sum_{j=i}^{N_{\alpha}} \frac{sin(qr_{ij})}{qr_{ij}}
\end{equation}
Where $N_{\alpha}$ are the number of species within a distance cutoff $r^{\alpha}_{cut}$. Taking inspiration from Ornstein–Zernike\cite{ornstein1914accidental}, Kirkwood\cite{kirkwood_statistical_1935} equations and others\cite{van_leeuwen_new_1959, salpeter_mayers_1958}: the final local symmetric order centering Pb positions \textit{$r_i^{Pb}$} is calculated by multiplying these individual local contributions\cite{ahlawat_atomistic_2020} from different species:
\begin{equation}\label{eqn:CV3}
 \xi_i=\prod_{\alpha = Cs, Pb, I} s^{\alpha}(r_i^{Pb})
\end{equation}
The final expression for the CV becomes the sum for all Pb species:
\begin{equation}\label{eqn:CV4}
 S_{p}= \sum_{i=1}^{N_{Pb}} \xi_i
\end{equation}

Using $S_{p}$, we conduct biased simulations employing a reduced charge potential of CsPbI\textsubscript{3}\cite{bischak_liquid-like_2020}. We successfully observe the crystallization of perovskite structure, presented in Figure \ref{fig:full} and supplementary movie SM1. Additional simulation details can be found in the Methods section. Our observations reveal that the solid-solid phase transition is a complex, multi-step process, generating various stack-faulted structures, anti-phase domains, and $<110>$-edge dislocations\cite{hirel_glissile_2016, song_direct_2022} during the transition from full edge-sharing to corner-sharing structures of CsPbI\textsubscript{3}. \\

\onecolumngrid

\begin{figure}
  \includegraphics[width=1.0\linewidth]{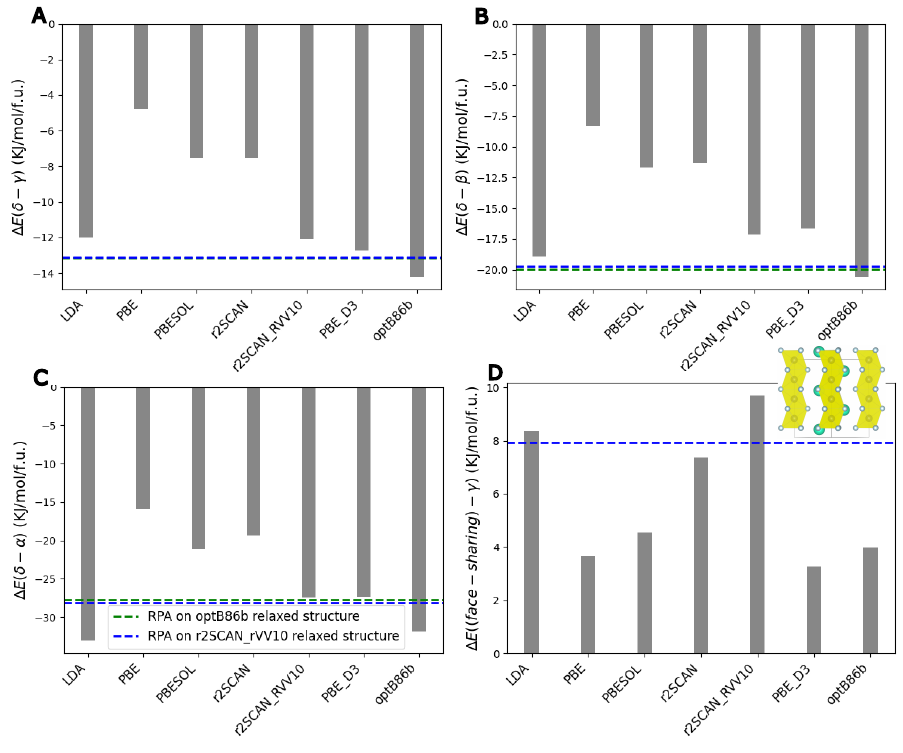}
  \caption{\label{fig:rpa_dft} \textbf{Relative Energies of different polymorphs using RPA+HF and a range of DFT functionals}}
\end{figure}

\twocolumngrid

To further scrutinize these observations, we employ an alternative CV based on permutation invariant vector (PIV) distances\cite{gallet_structural_2013, pipolo_navigating_2017}. Initially, PIV distances are calculated for the starting configuration concerning the reference structures of the edge-sharing ($\delta$) and corner-sharing ($\alpha$) phases of CsPbI\textsubscript{3}. Subsequently, we utilize these indices to define the functional form of the CV as follows: 

\begin{equation}\label{eqn:CV5}
 S_{PIV}= f(r|r_\alpha)/(f(r|r_\alpha) + f(r|r_\delta))
\end{equation}

Where $f(r|r_\alpha)$ is the PIV distance from the corner-sharing phase, $f(r|r_\delta)$ is the PIV distance from edge-sharing phase, $r$ are the coordinates of the system, {$r_\alpha$} are the coordinates of the reference $\alpha$-phase phase and {$r_\delta$} are the coordinates of the reference $\delta$-phase. Using this functional form in Equation 5, we perform biased simulations, and again successfully observe a similar solid-solid phase transition pathway in CsPbI\textsubscript{3}. \\

Motivating from these exciting observations, we further investigate direct solid-solid phase transitions with the machine learning potentials (MLPs). In this work, Nequip MLPs\cite{batzner_e3-equivariant_2022} are built from dataset of forces, energies and stresses acquired by performing DFT driven MD simulations of CsPbI\textsubscript{3} in the isothermal–isobaric ensemble. \\

However, the accuracy of DFT calculations may depend on the exact exchange-correlation functional (XC) adopted in the procedure. Therefore, we first conduct an elementary comparison of widely employed XC functionals against reference \textit{first-principles} many-body electronic structure method of random phase approximations (RPA+HF)\cite{bohm_collective_1951, gell-mann_correlation_1957, harl_assessing_2010, kaltak_cubic_2014, braeckevelt_accurately_2022} which can account for dynamic electronic screening and long-range dispersion interactions. Figures \ref{fig:rpa_dft}A-C show the relative ground state energies of different perovskite polymorphs against the thermodynamically stable $\delta$-phase of CsPbI\textsubscript{3}. The best results are obtained for the PBE+D3(BJ)\cite{becke_density-functional_2005, johnson_post-hartree-fock_2006, grimme_consistent_2010, grimme_effect_2011},  r2SCAN+rVV10\cite{ning_workhorse_2022}, optB86b\cite{klimes_van_2011} functionals that include van der Waals interactions. We note that although acceptable relative energies are obtained with LDA\cite{hohenberg_inhomogeneous_1964, kohn_self-consistent_1965}, however this might be due to error cancellations where exchange energy is generally underestimated, and the binding energy is overestimated. To choose the best, we perform an additional calculation of ground state relative energies between $\gamma$-phase and manually constructed polymorph of CsPbI\textsubscript{3} based on the framework of Pb-I face-sharing octahedra\cite{ahlawat_crystallization_2023}. This structure is prepared by replacing formamidnium cations with cesium into hexagonal phase of formamidnium lead iodide (FAPbI\textsubscript{3}), illustrated in Figure 3D. Numerous experimental investigations\cite{lee_formamidinium_2015, yi_entropic_2016, merten_quantifying_2021} have consistently demonstrated that the introduction of cesium into hexagonal-FAPbI\textsubscript{3} rapidly form the black perovskite phase, and prevent the formation of its hexagonal face-sharing counterpart. \\

\begin{figure}
 \centering
  \includegraphics[width=0.80\linewidth]{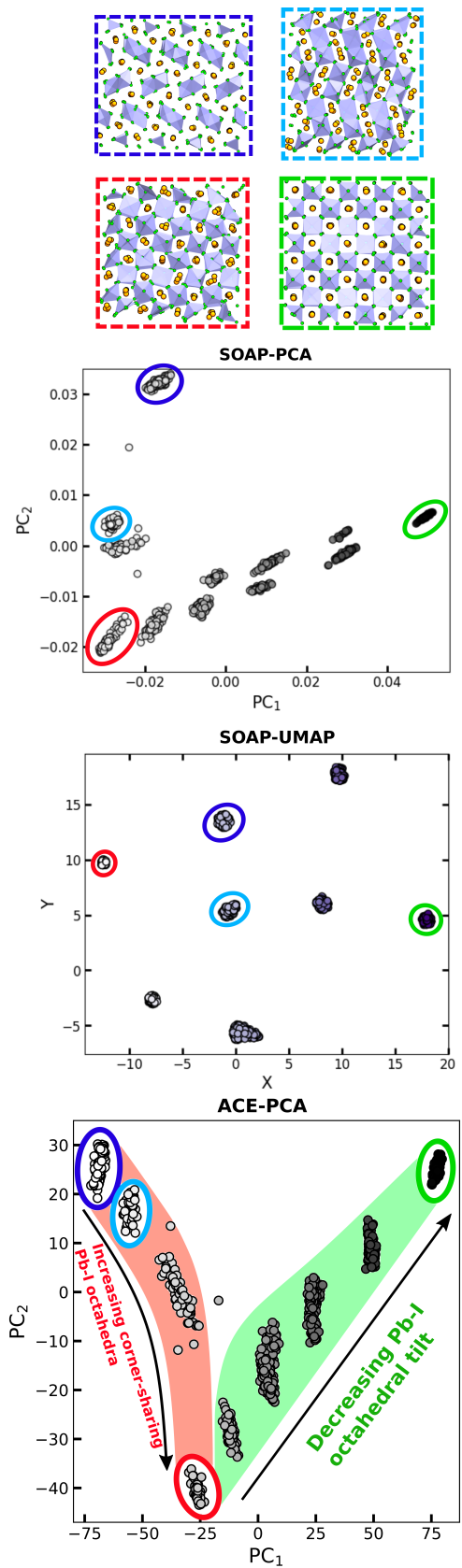}
  \caption{\label{fig:umap} \textbf{Classifications of trajectories with SOAP and ACE}}
\end{figure}

Consequently, one anticipates that a face-sharing structure doped with cesium would exhibit instability and higher energy levels compared to the $\gamma$-perovskite phase, a phenomenon corroborated by the ground state energies presented in Figure \ref{fig:rpa_dft}D. Taking into account the comprehensive set of results presented in Figure \ref{fig:rpa_dft}, we opt for the utilization of the \textbf{r2SCAN+rVV10} functional, considering its suitable depiction of relative energies across various structural configurations. It is worth noting that Shuxia Tao and co-workers\cite{xue_first-principles_2021, xue_compound_2023} have also previously utilized SCAN+rVV10\cite{peng_versatile_2016} to investigate point defects in various perovskites. Subsequently, we perform DFT based \textit{ab-initio} MD (AIMD) simulations\cite{payne_iterative_1992, carignano_critical_2017} using the selected functional. Nequip MLPs are generated based on the dataset acquired from the AIMD simulations. The Methods section provide details of these simulations. Using MLPs, we again perform a series of biased simulations to further scrutinize the phase transformation dynamics. Remarkably, the observed pathway is similar to the transformation sequence previously portrayed in Figure \ref{fig:all}. This noteworthy consistency between the outcomes obtained through distinct simulation methodologies, RCPs and MLPs underscores the robustness and reliability of our findings. \\

To further understand the structural transformations, next we perform a detailed analysis of simulated trajectories using smooth overlapping of atomic positions (SOAP)\cite{bartok2010gaussian, musil_machine_2018, musil_efficient_2021} and atomic cluster expansion (ACE) \cite{drautz2019atomic,dusson_atomic_2022,witt_acepotentials.jl:_2023} descriptors. \\

\begin{figure}
  \includegraphics[width=1.0\linewidth]{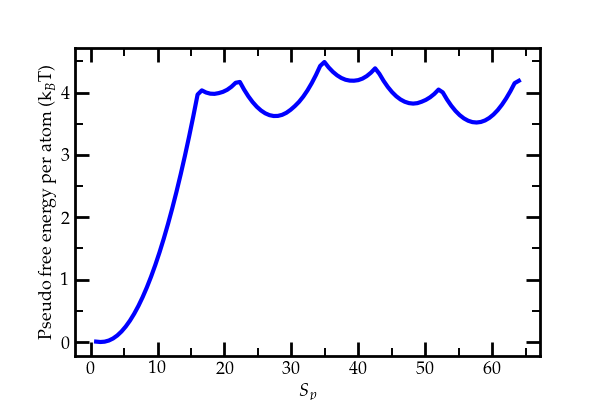}
  \caption{\label{fig:sfes}Qualitative free energy profile for direct solid-solid phase transition.}
\end{figure}

SOAP and ACE, known for capturing detailed atomic variations, are complemented by principal component analysis (PCA) and uniform manifold approximation and projection (UMAP)\cite{mcinnes_umap:_2018} for dimensionality reduction. This combined approach enhances our ability to comprehend intricate molecular relationships and dynamics. The resulting Figure 4 represents our simulated trajectories, highlighting clusters of similar structural conformations and uncovering underlying patterns in the simulated trajectories. Beyond the discerned mixtures of corner-sharing and edge-sharing structures highlighted in Figure \ref{fig:umap}, our findings suggest the potential existence of distinct tilted perovskite configurations (see clusters that are not circled and highlighted in Figure \ref{fig:umap}) on the free energy surface, see Figure \ref{fig:sfes}. It is noteworthy to acknowledge the prior simulations conducted by Wiktor, Erhart and co-workers, have employed Glazer notations to rationalize the presence of diverse locally tilted octahedra during tilt-based phased transformations in cesium lead bromide\cite{wiktor_quantifying_2023}. Whereas, we show that the wide range of tilted-perovskite phases could exist on the free energy surface during solid-solid crystallization and can be quantified using machine learning techniques. Our observations broadens the understanding of the diverse structural landscapes in halide perovskites and can be easily extended to wide range of perovskite materials, in particular in oxide perovskites where tilt-based phase transformation play a crucial role in determining their electronic properties for designing electronic devices. \\

Nonetheless, to achieve a realistic material simulation of phase transition dynamics, it is important to extend beyond the collective atomic motions of approximately 1000s of atoms. It is crucial for capturing the detailed macroscopic mechanisms and dynamics of solid-solid crystallization on a larger scale ensuring a direct atomic visualization of experimental processes, in contrast to small-scale simulations. This transition sets the stage for the subsequent section, wherein we try to delve into the investigation of the macroscopic aspects of solid-solid phase transitions. \\

\begin{figure}[hbt!]
  \includegraphics[width=1.01\linewidth]{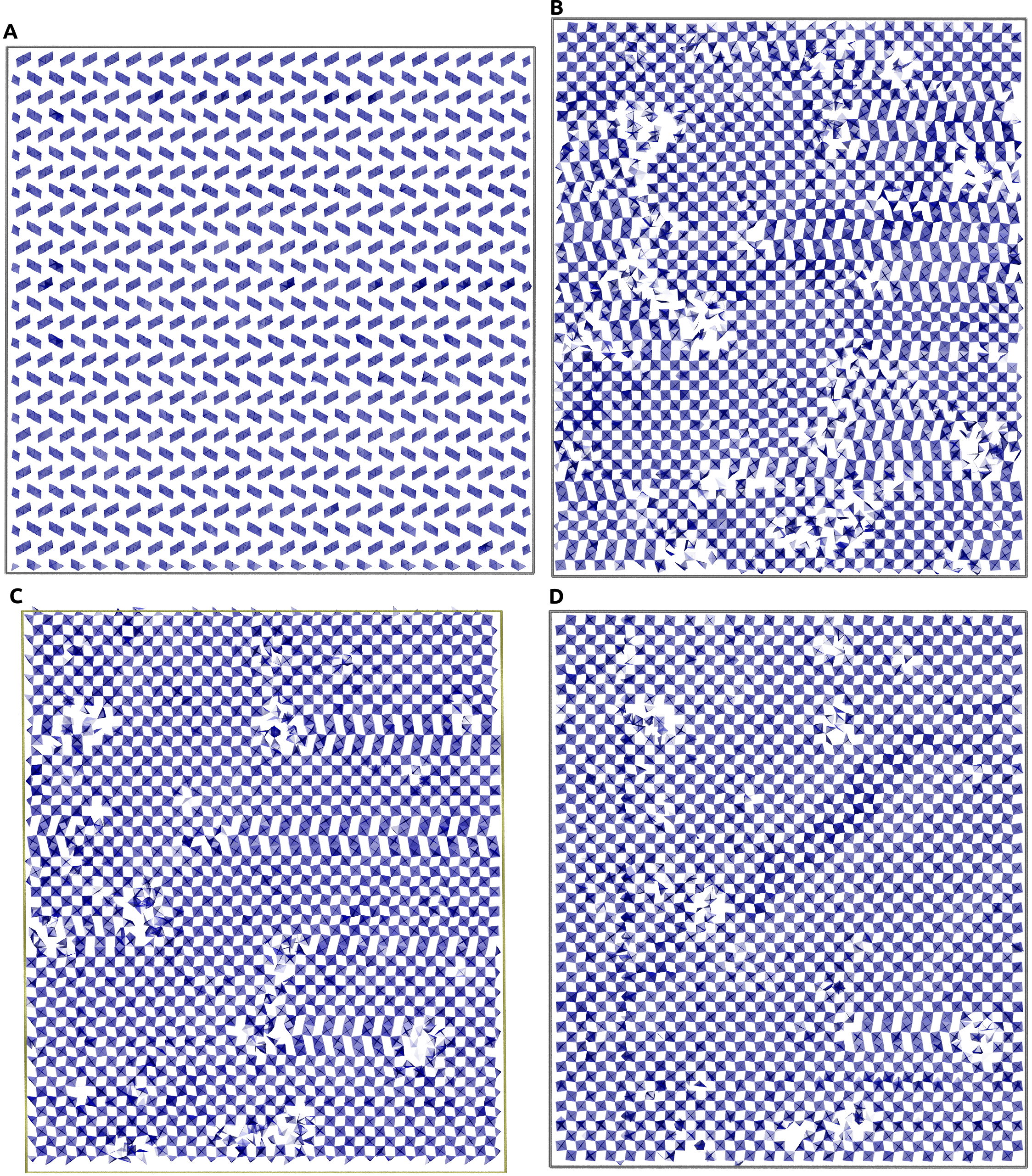}
  \caption{\label{fig:l2} \textbf{Solid-solid phase transition by enhanced sampling}: Figure shows the simulated transformation of edge-sharing $\delta$-phase (A) to corner-sharing perovskite-phase of CsPbI\textsubscript{3}. Selectively Pb-I structures are represented as blue octahedra to guide the eye.}
\end{figure}

To comprehend the macroscopic dynamics of solid-solid phase transitions, we initiated biased simulations on relatively large-scale structures comprising up to 35,000 atoms. As illustrated in Figure 2, our observations reveal that the phase transition progresses through the sliding of octahedra. To emulate this on large scales, we adopt an initial configuration of the $\delta$-phase, characterized by the relatively larger dimensions along the X and Z directions, however relatively smaller dimensions along the Y-direction, approximately measuring 250 × 24 × 250 \AA, illustrated in Figure \ref{fig:l2}A. Once again, using enhanced sampling simulations with a multi-species structure factor as CV, we excitingly witness the emergence of corner-sharing perovskite structures, illustrated sequentially in Figures \ref{fig:l2}A-B-C-D. Upon analyzing the trajectory data and Figure \ref{fig:l2}B, we again observe that the morphological evolution occurs through the mixed formation of stack-faulted or edge-dislocated configurations. This phenomenon is directly observable in Figures \ref{fig:l2}B-D, where significant amount of layers featuring mixed edge-sharing and corner-sharing arrangements are crystallized during the phase transition. Supplementary movie SM2 provides a detailed atomic-scale depiction of this phase transformation. \\

However, a crucial insight emerges from our large-scale simulations. As demonstrated in Figure \ref{fig:l2}D, the process gives rise to several defects, particularly observed in the neighbouring regions where mixed-corner-edge-sharing layers have crystallized. Despite the subsequent conversion of these mixed configurations to full corner-sharing perovskite structures during the biased simulations, the formation of defects is evident, and defects prevailed in the final configuration. These are easily recognizable by incomplete or broken octahedra (and empty spaces) in Figure \ref{fig:l2}D. In opposite to that, where parts of the $\delta$-phase that undergo direct conversion to the perovskites instead of mixed corner-edge-sharing structures, exhibit fewer defects, and can be readily visualized in Figures \ref{fig:l2}A, B and Supplementary movie SM2. From this comprehensive analysis, we propose that one of the mechanism of underlying defect formation in halide perovskites may strongly involve the formation of long-range intermediate structures during the crystallization process. \\

Simulating structural transformations arising from biased simulations involving more than 35,000 atoms surpasses the capabilities of current code implementations and available computational resources. As a result, in the following section, we undertake brute force MD simulations to investigate various aspects of solid-solid crystallization in greater depth.

\newpage
\onecolumngrid

\section{\label{sec:level1}Large scale seeded crystallization\protect\\ }

\begin{figure} [hbt!]
  \includegraphics[width=1.0\linewidth]{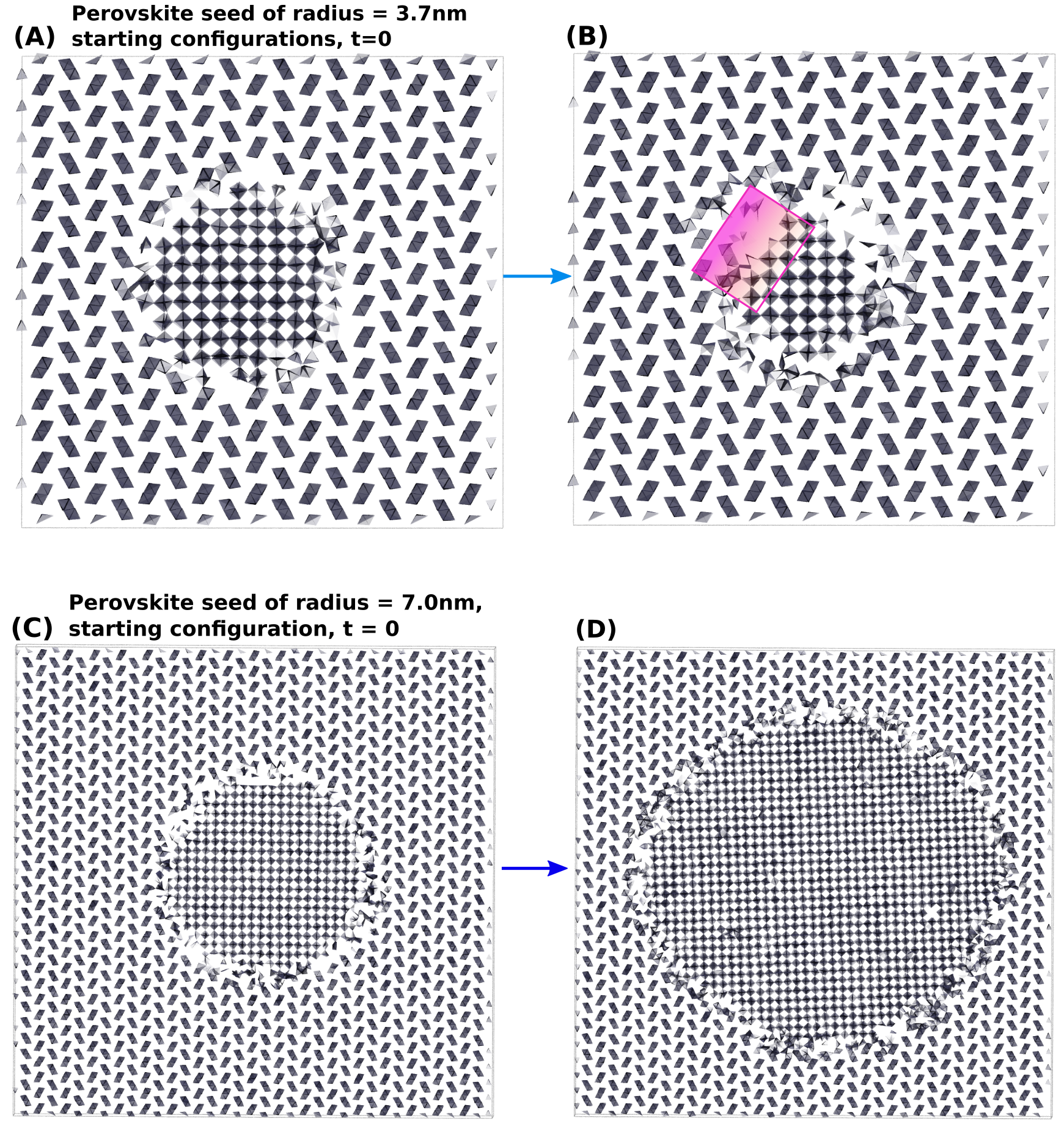}
  \caption{\label{fig:seed} \textbf{Seeded simulations:} (A) Snapshot depicting the starting configuration with a seed size of 3.7 nm, and (B) illustrates its time evolved configuration. (C) Presents a snapshot of the initial structure with a seed size of 7.0 nm, while (D) showcases the evolved structure. To enhance clarity, middle slices from the 3D box's center are shown here to guide the eye. The representation is focused on Pb-I octahedra for clear distinction between the two phases.}
\end{figure}

\newpage
\twocolumngrid

Earlier experiments\cite{chen_size_1997, jacobs_activation_2001, li_revealing_2021, peng_two-step_2015, peng_situ_2023} and simulations\cite{santos-florez_size-dependent_2022, ball_size_2022, khaliullin_nucleation_2011} investigating solid-solid phase transitions in materials other than perovskites, have noticed that following the formation of initial smaller nuclei in a foreign crystalline solid, the subsequent solid-solid crystallization process may obey classical nucleation pathways\cite{bolhuis_isostructural_1994, debenedetti_metastable_1996, auer_prediction_2001, bai_test_2005, sanz_homogeneous_2013} on larger length scales. Seeking a fundamental macroscopic level understanding in perovskite, we perform \textbf{unbiased brute force seeded simulations}\cite{ panagiotopoulos_direct_1987, bai_calculation_2006, gonzalez-salgado_solidsolid_2011, espinosa_seeding_2016, espinosa_homogeneous_2014, espinosa_fluid-solid_2013, binder_phase_2021}, a sophisticated approach that could allows us to explore and probe underlying mechanisms governing macroscopic solid-solid phase transitions in halide perovskites. \\

\begin{figure}
  \includegraphics[width=1.0\linewidth]{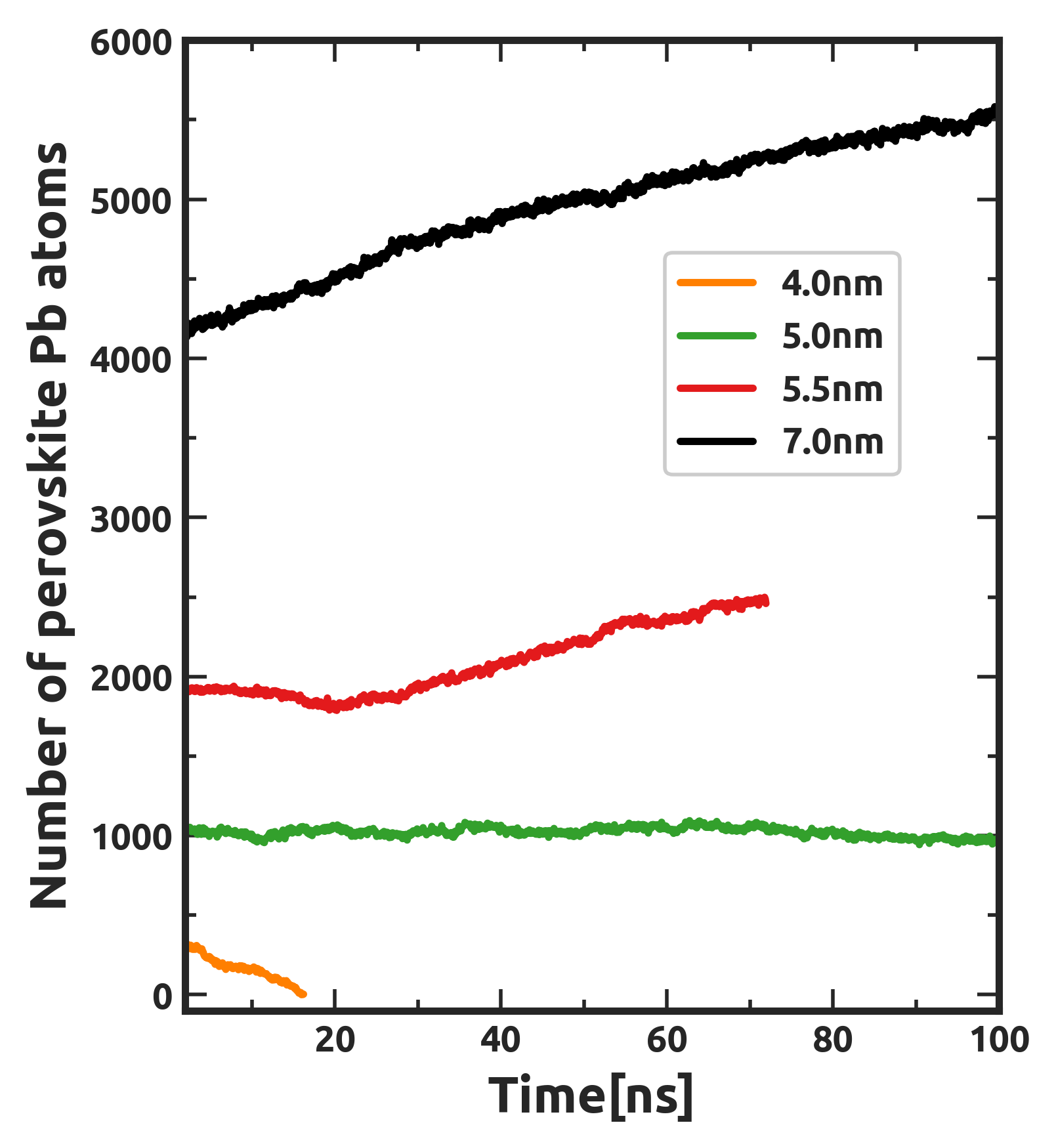}
  \caption{\label{fig:seed_q4} \textbf{Seeded simulations:} Figure shows the time evolution of amount of perovskite phase calculated using lq4.}
\end{figure}

To start with, we prepare different structural configurations by incorporating spherical structures (seeds) of perovskite phase into extensive yellow-phase structures, details are provided in the Methods section. We select perovskite seeds with varying radii ranging from 3.0 to 7.0 nm which could offers a comprehensive exploration of nucleation and growth processes at different initial conditions. Starting from equilibrated configurations with different initial seeds we carry out brute force MD simulations using RCPs and monitor the time evolution of fraction of perovskite in these systems. To assess the quantity of perovskite within our simulations, two distinct metrics are employed. Firstly, the relative intensity of the corresponding (100) X-ray peak of perovskite-phase is utilized, allowing for tracking and comparison with experimental data. Additionally, a local-bond order parameter lq4\cite{rein_ten_wolde_numerical_1996, van_duijneveldt_computer_1992, lechner_accurate_2008}, is employed to precisely quantify the number of formula units corresponding to different phases within the simulation box. To compute the X-ray peaks for these extensive structures, we employ a simplified Debye form of structure factor, more simulation details are provided in the Methods section. \\

\begin{figure}
  \includegraphics[width=0.95\linewidth]{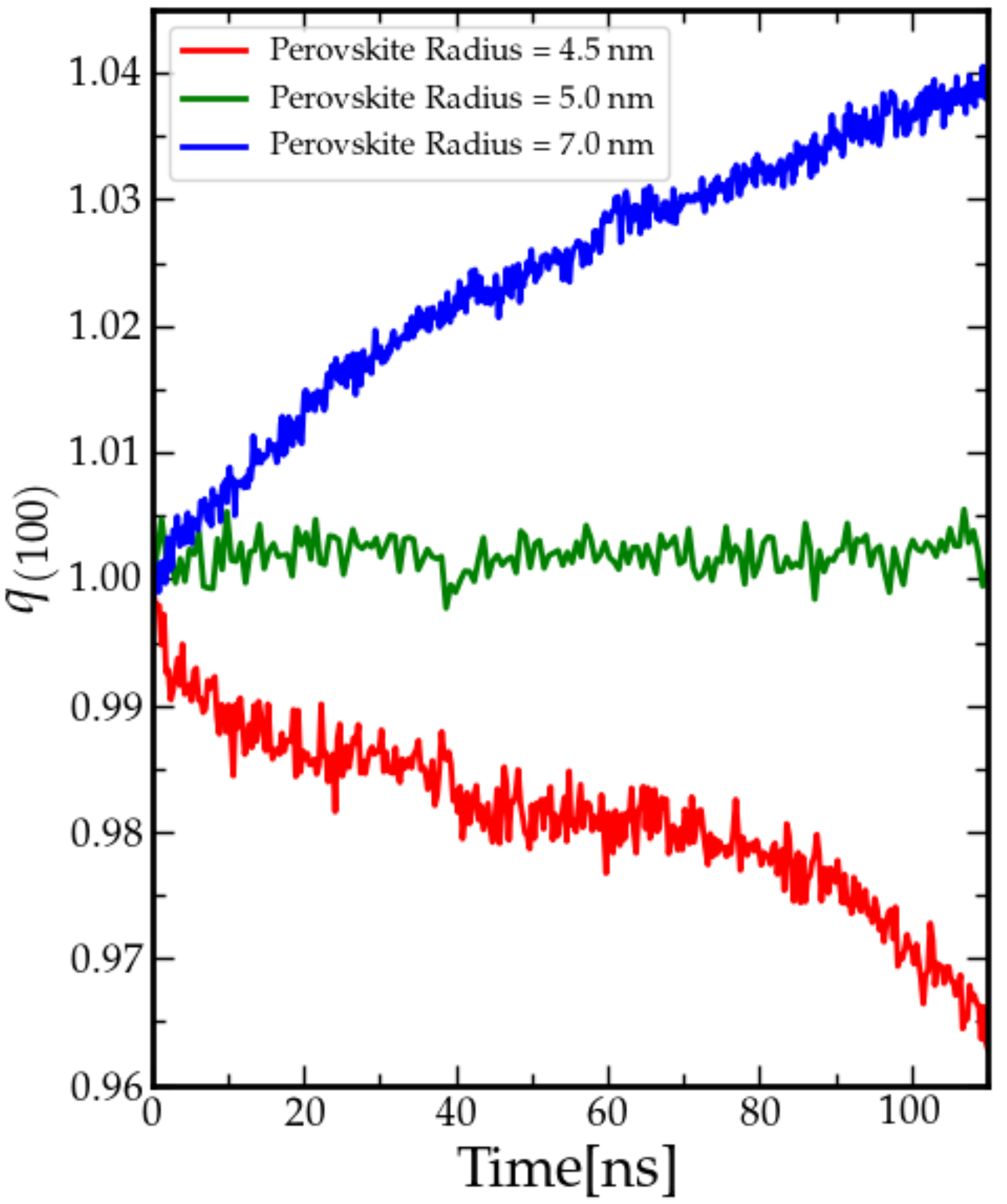}
  \caption{\label{fig:seed_sf} \textbf{Seeded simulations:} Figure shows the time evolution of amount of perovskite phase calculated using structure factor.}
\end{figure}

From simulations, we observe that as the fraction of perovskite decreases, the inserted seed becomes subcritical and completely transform into $\delta$-phase. Conversely, as the fraction increases, the cluster transitions to a post-critical state. This approach allows us to ascertain whether the inserted cluster is critical, similar to the methodologies employed in earlier seeded simulations involving ice\cite{sanz_homogeneous_2013} and carbon-graphite-diamond transitions\cite{khaliullin_nucleation_2011}. Figures \ref{fig:seed}(A) and \ref{fig:seed}(C) depict the initial configurations for seed sizes of 3.7 nm and 7.0 nm, respectively, serving as the starting points for our simulations. Subsequently, Figures \ref{fig:seed}(B) and \ref{fig:seed}(D) show the concluding configurations for the respective seeded configurations. First, the visual analysis of these figures directly reveals that the smaller perovskite seed experiences significant contraction, as illustrated in Figure \ref{fig:seed}(B), ultimately transforming into the $\delta$-phase characterized by an increase in edge-sharing structure. \\

\newpage

\begin{figure}
  \includegraphics[width=0.98\linewidth]{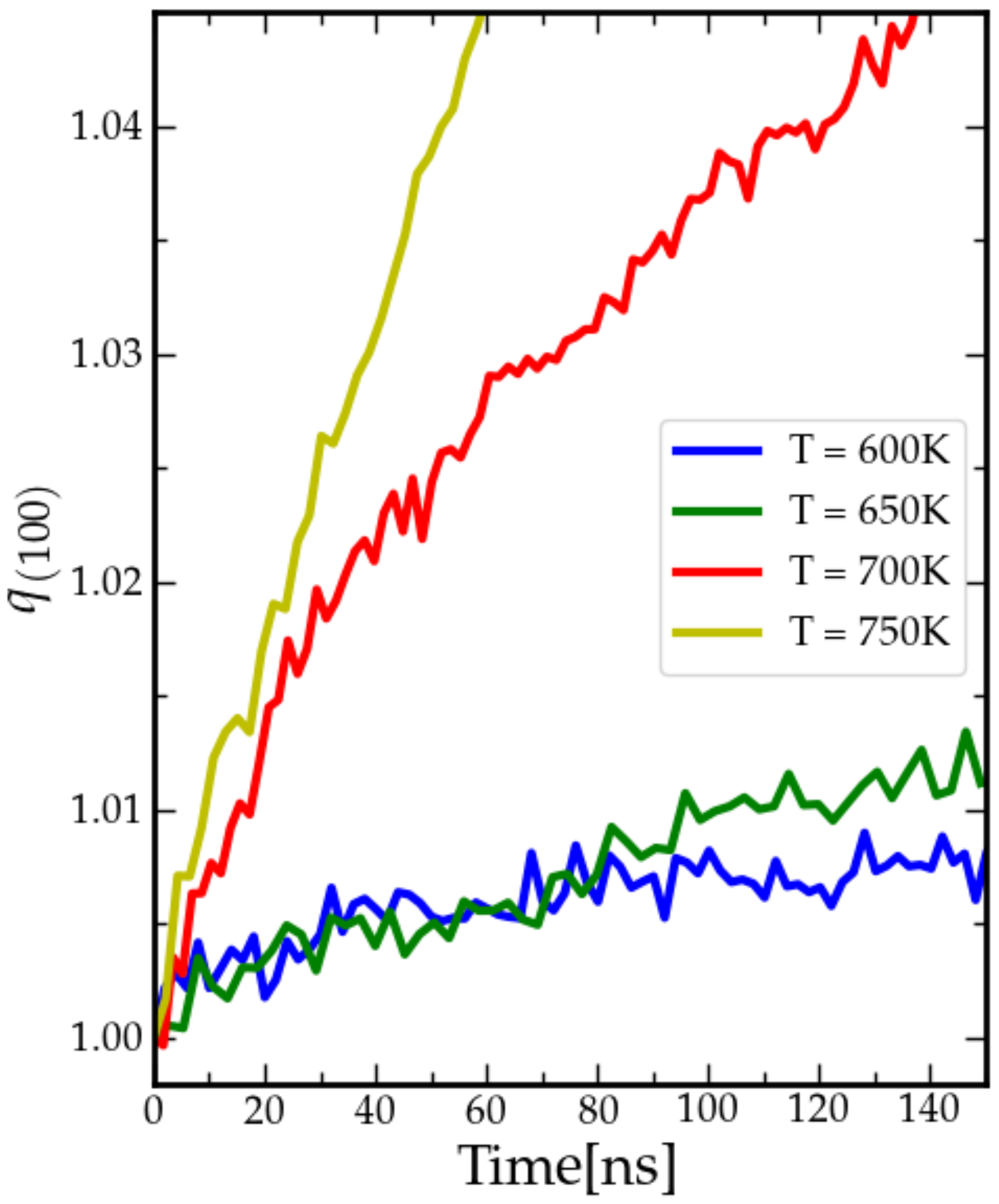}
  \caption{\label{fig:seed_sfg} \textbf{Seeded simulations with 7 nm seed:} Figure illustrates the time evolution of the normalized (100)-peak in the structure factor at different temperatures for a seed radius of 7.0 nm.}
\end{figure}

Conversely, the larger seed (7.0 nm) exhibits substantial growth, resulting in the formation of a very larger perovskite crystal structure. Figures \ref{fig:seed_q4} and \ref{fig:seed_sf} illustrate the perovskite fraction over time for various simulated perovskite seeds. \\

Interestingly, a noteworthy observation arises at the boundary between the two phases, where the atoms exhibit liquid-like behavior. This underscores a dynamic moving interface between the coexisting phases, suggesting that solid-solid phase transition is facilitated by the propagation of a liquid-like front\cite{pogatscher_solidsolid_2016, peng_two-step_2015, sanz_homogeneous_2013, bischak_liquid-like_2020}. The interplay between liquid-like behavior and crystallization at the phase boundary adds a layer of complexity to the understanding of solid-solid crystallization in perovskites. \\

To complement with visual observations, simulation movies SM3 and SM4 are provided in the SI for the dynamical insights into the temporal evolution of these systems. Importantly, it is crucial to note that these simulations are conducted at significantly lower temperatures of 600, 650 and 700K than the calculated melting temperature 800K of CsPbI\textsubscript{3}, see the Methods section for details. This deliberate choice ensures the liquid-like front does not results from the melting of the crystal. \\

\begin{figure}
  \includegraphics[width=0.99\linewidth]{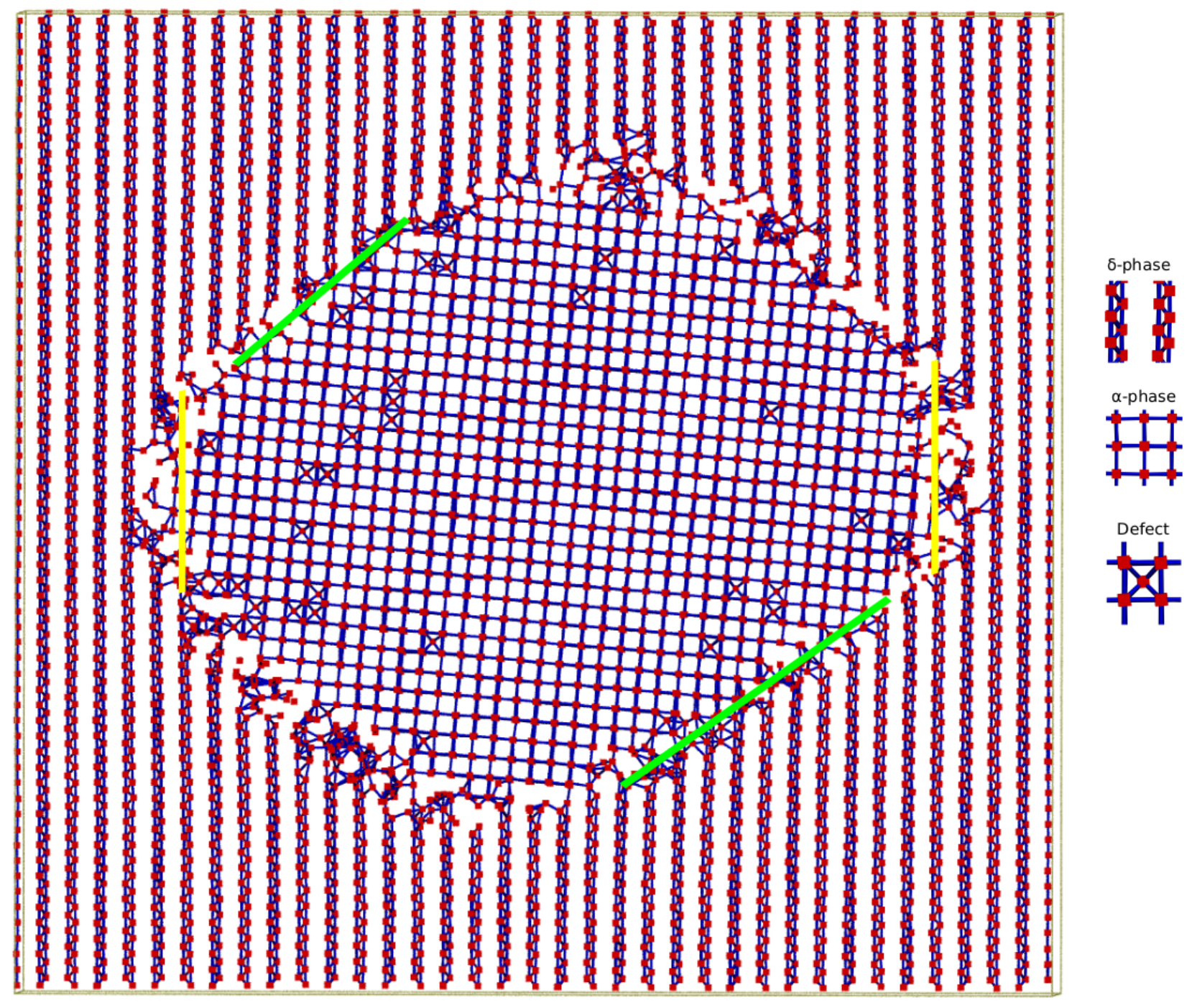}
  \caption{\label{fig:ls-crys} \textbf{Seeded Simulations}. Figure displays a configuration of the ultimately crystallized structure. For clarity, only a slice from the center of the seed is presented, featuring connected Pb atoms exclusively to guide the eye.}
\end{figure}

In the following sections, we delve into the mechanistic details of the solid-solid nucleation mechanism. Figure \ref{fig:seed_sfg} offers an insightful perspective into the growth profile of a larger seed across different temperatures. The impact of temperature on growth rate is evident, with a pronounced acceleration at higher temperatures beyond 650K, similar to the earlier reported experiments\cite{chen_efficient_2021}. However, this heightened growth rate is simultaneously accompanied by the emergence of defects within the crystal structure. Subsequently, we scrutinize these configurations and find the formation of different type of point defects, as portrayed in Figure \ref{fig:ls-crys}. For clarity, we selectively depict connected Pb atoms, highlighting displaced or defected Pb atoms with cross marks on the evolving perovskite lattice. \\

Furthermore, we conduct room-temperature relaxation of these final structures at first-principle level of calculations. Performing direct \textit{ab-initio} simulations for systems of this magnitude is utterly unfeasible. We achieve this by employing MLPs through the neuroevolution potential (NEP) at r2SCAN+rVV10 level and using GPUs accelerate MD (GPUMD) code for molecular dynamics simulations of these large scale configurations. Detailed procedures are explained in the Methods section. These structurally relaxed configurations can serve as valuable input for the comparisons with transmission electron microscopy (TEM) experiments and spectroscopic experiments, thereby offering direct visualizations for tracking the phase transition, formation of defects and help experiments to obtain atomic-level details of the heterogeneities during crystallization process. \\

Moreover, a significant finding in our investigation is related to the dynamic evolution of the perovskite phase. We discern a distinct discrepancy in the evolution rates of different facets of the perovskite phase. Specifically, the (100)-facet displays accelerated growth compared to its counterparts, as evident in Figure \ref{fig:seed} and Figure \ref{fig:ls-crys}. These figures illustrates the rapid expansion of the (100)-facet boundary, originating from an initial spherical seed. However, when considering the boundary  of the (110)-facet and the $\delta$-phase, interestingly, we observe the formation of a coherent phase between them. Figure \ref{fig:ls-crys} provides direct evidence of this phenomenon, where the green color interface highlights the (110)-facet of the grown crystal, while the yellow color interface represents the (100)-facets. \\





\newpage

\onecolumngrid

\twocolumngrid


\begin{figure}
  \includegraphics[width=\linewidth]{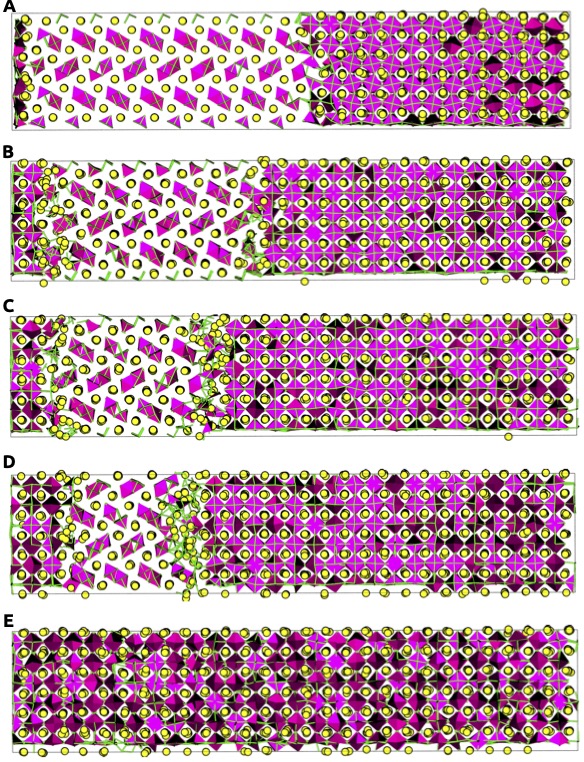}
  \caption{\label{fig:co-solid} \textbf{Solid-solid co-existing simulations of (100)$_{\delta}\|(100)_{\alpha}$}: Figures A to E illustrate the temporal progression of structural changes during co-existing simulations. For clarity, only magenta-colored Pb-I octahedra are displayed alongside yellow cesium atoms. These visualizations were generated using VMD software\cite{humphrey_vmd:_1996}.}
\end{figure}

\begin{figure}
  \includegraphics[width=0.99\linewidth]{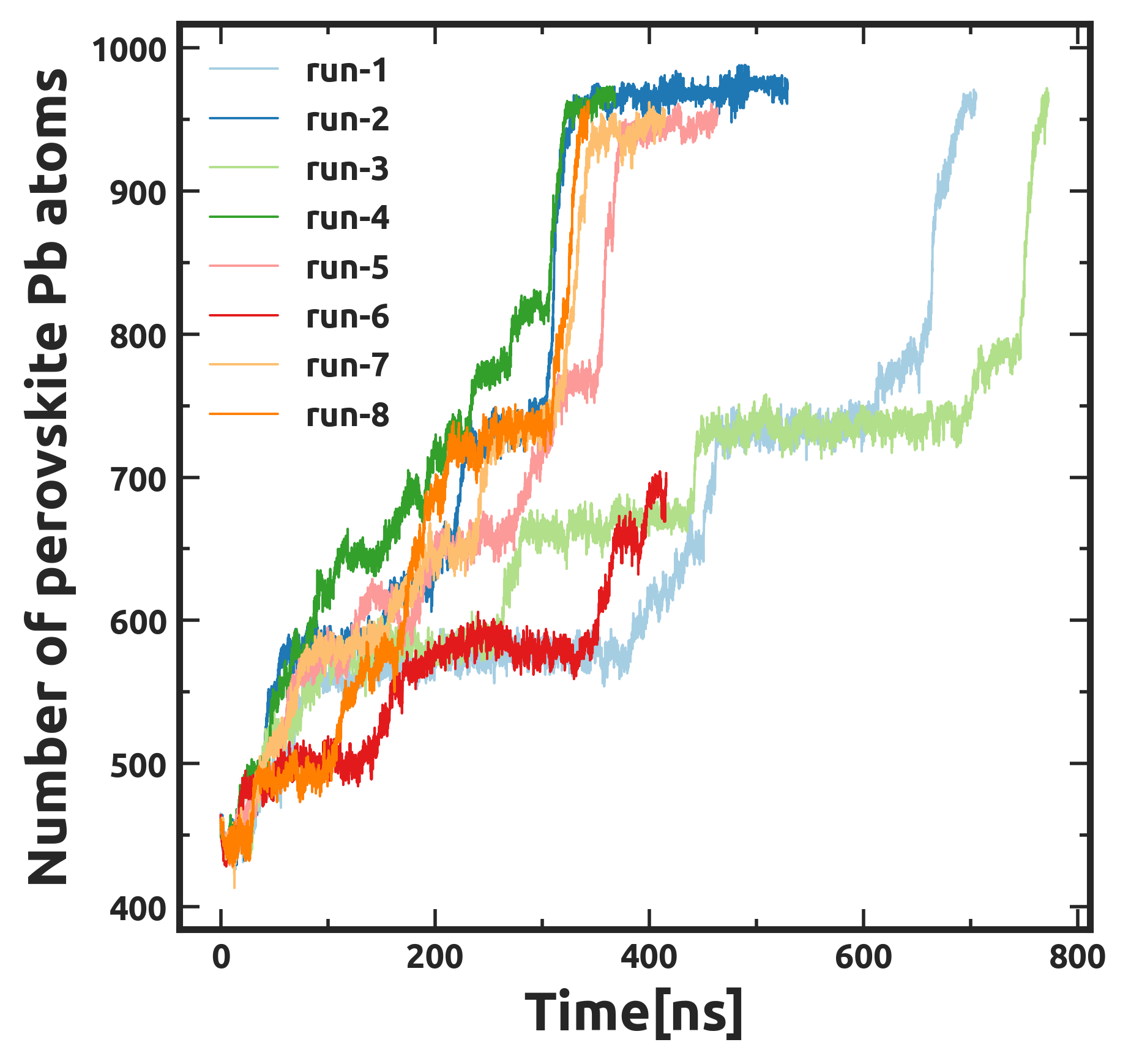}
  \caption{\label{fig:co_colvar} \textbf{Time evolution of local bond order parameter (lq4$>$0.5) for solid-solid co-existing simulations of (100)$_{\delta}\|(100)_{\alpha}$}}
\end{figure}

\begin{figure}
  \includegraphics[width=0.99\linewidth]{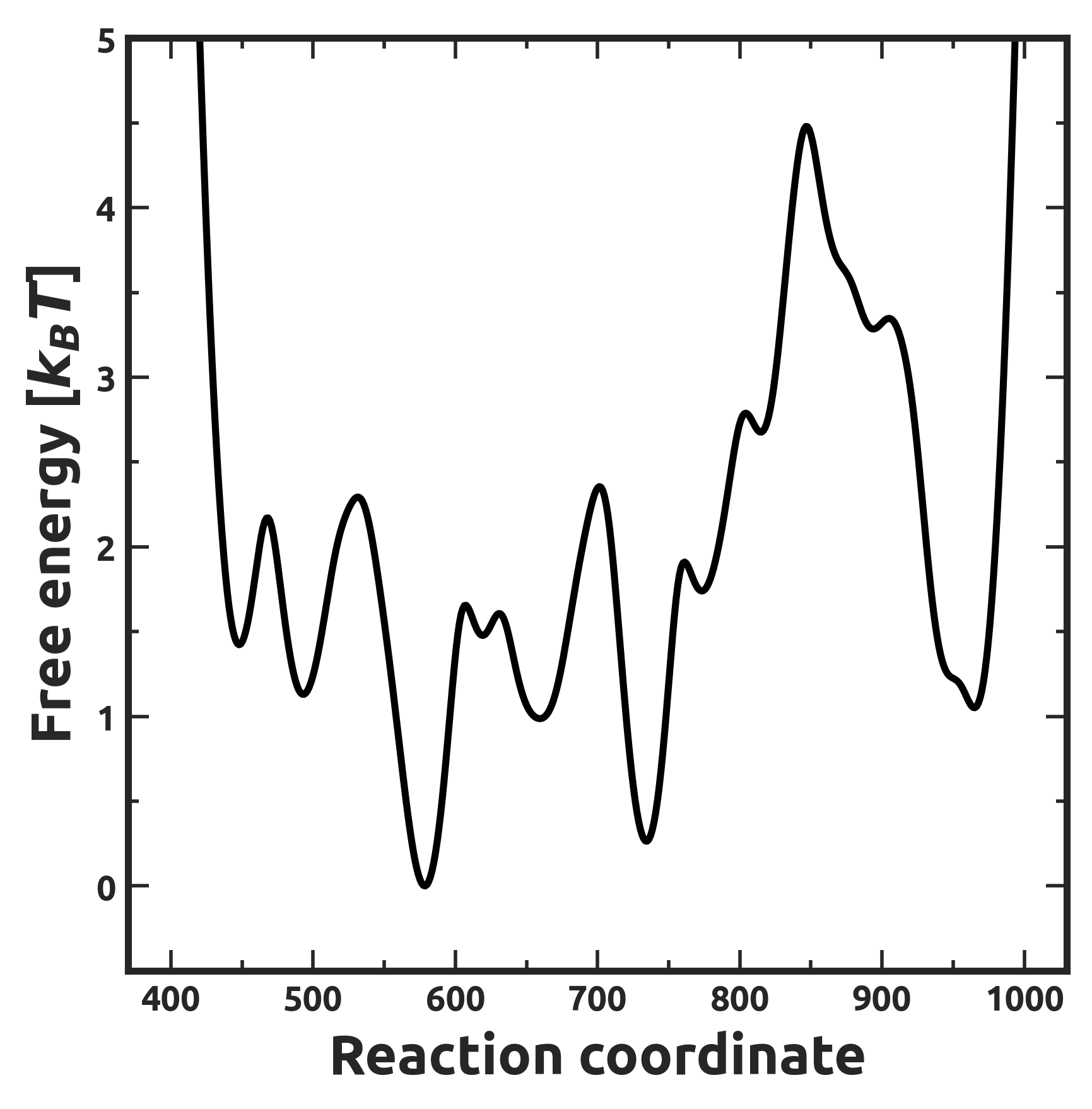}
  \caption{\label{fig:co_fes} \textbf{Free energy profile of the solid-solid co-existing simulations of (100)$_{\delta}\|(100)_{\alpha}$}}
\end{figure}

To further investigate this growth process of respective (100)-perovskite-facet, we perform solid-solid co-existing simulations of utilizing NEP MLPs and GPUMD code. Detailed methodologies and simulation set-ups are provided in the Methods section. Brute force MD simulations are conducted at temperatures of 600K, 650K, and 700K, revealing a \textbf{layer-by-layer transformation}\cite{ahlawat_crystallization_2023} of the $\delta$-phase into the perovskite phase at different temperatures. This whole transformation is illustrated in Figure \ref{fig:co-solid} and Supplementary Movie SM5. Further characterization of the layer-by-layer transformation is evaluated by calculating the change in the number of Pb species belonging to the perovskite phase using the previously defined lq4. Figure \ref{fig:co_colvar} shows the time evolution of the perovskite phase across eight repeated MD runs started with the variable momenta. The step-increase in the number of perovskite Pb atoms indicates that upon the onset of perovskite crystallization at the co-existing interface, a complete layer of the $\delta$-phase transitions into the perovskite phase. Additionally, we computed the free energy profile of this process using the equation F = k$_B$T log(p($lq4$)). The free energy profile of the co-existing simulations at T=700K is depicted in Figure \ref{fig:co_fes} and eventually indicate a layer-by-layer nature of the transformation, as evidenced by the individual free energy minima separated by barriers on the order of k$_B$T.  \\

Now, we investigate scenarios involving sub-critical nuclei. In such instances, we observe a distinctive behavior in which the (110)-facet of the perovskite seed undergoes conversion to the [010]-facet of the $\delta$-phase at an accelerated rate compared to other facets, as highlighted in Figure \ref{fig:seed} B. Furthermore, we also perform co-existing simulations of the (110)-perovskite facet with the $\delta$-phase at 700 K are aimed to understand the interface dynamics. Interestingly, seven time repeated simulations lasting up to 2 microseconds did not show any significant increase in the perovskite structure. This stability contrasts with the behavior observed for the (100) perovskite facet, where co-existence with the $\delta$-phase led to a complete conversion to perovskite structure by formation of a dynamical interface with a liquid-like character as shown in Figure \ref{fig:co-solid}C and D. In the case of the (110) facet, however, a distinct interfacial behavior emerged. Here, the (110) facet formed a well-defined and stable boundary with the $\delta$-phase, lacking the mobile characteristics of the boudary atoms observed with the (100) facet. This difference in interfacial dynamics between the two facets suggests a potential influence of crystallographic orientation on the stability and transformation behavior of the perovskite phase. \\

  
At the end, we try to analyze and compute the associated free energy barrier with the seeded crystallization process. We achieve this by gathering multiple histogram data derived from X-ray peak intensity across multiple seeded simulated runs, and combining them into a single run. \\

Utilizing the combined data, we calculate the free energy using the equation F = k$_B$T log(p($q_{100}$)), where p($q_{100}$) represents the probability distribution of the (100)-perovskite peak, as explained in prior discussions and detailed in the Methods section. Figure \ref{fig:seed_free} reveals a qualitative free energy barrier, highlighting the necessity to overcome an energy barrier on the order of 10 k$_B$T for solid-solid nucleation at 700K. Excitingly, our collective observations pave the way for quantifying the rates of solid-solid nucleation process at different conditions utilizing a modified classical nucleation theory framework\cite{chu_model_2000, li_revealing_2021}:

\begin{figure}
  \includegraphics[width=0.98\linewidth]{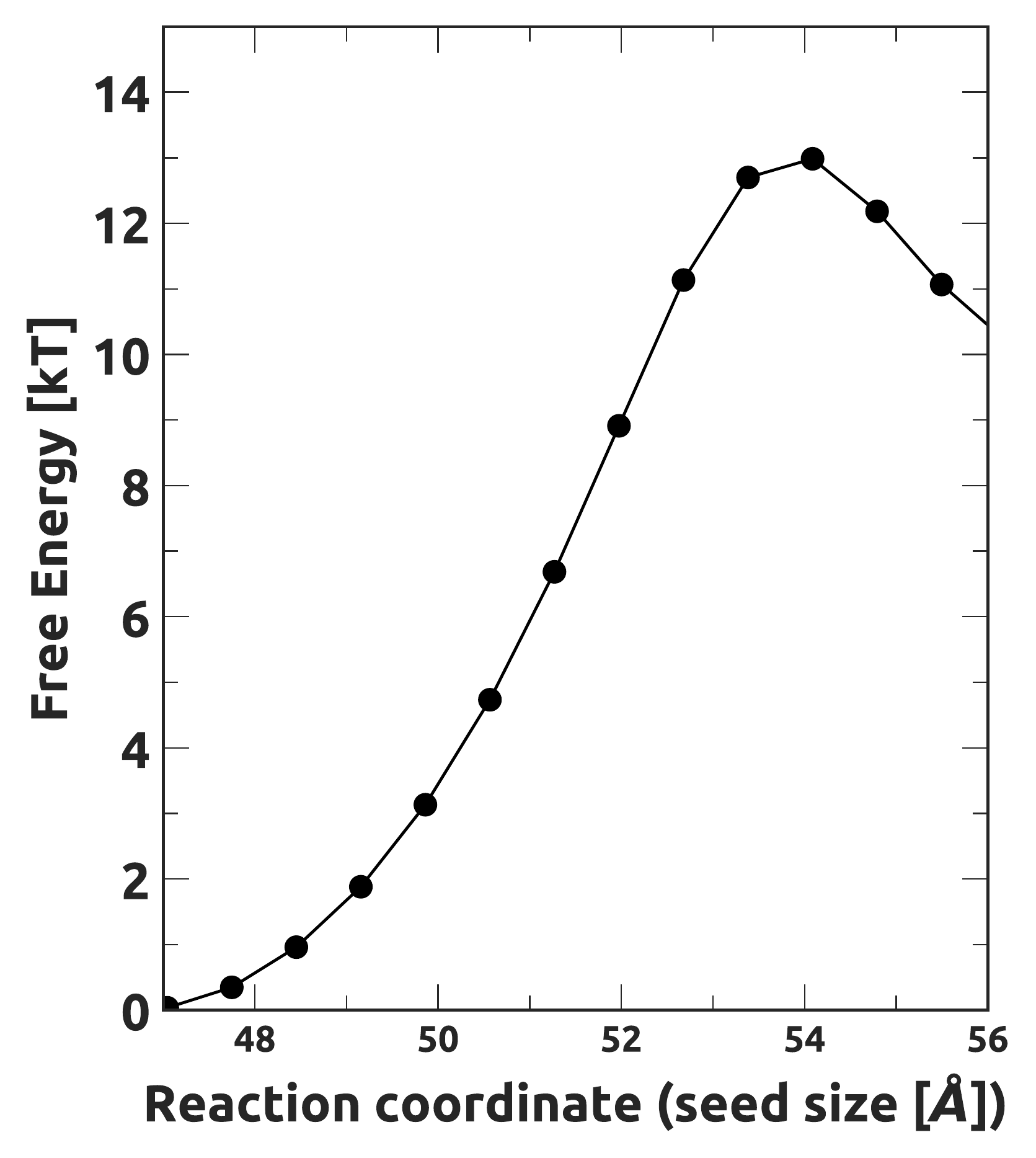}
  \caption{\label{fig:seed_free} \textbf{Seeded simulations:} Figure presents the qualitative free energy of crystallization across various seeded systems.}
\end{figure}

\begin{equation}\label{eqn:CV5}
 \Delta G = -V \rho \Delta \mu + S \gamma  + E_{strain} + E_{defect}
\end{equation}

Here, S represents the surface area, and V denotes the volume of the perovskite phase. Additionally, $\rho$ signifies the number density of the product nuclei, $\Delta \mu$ denotes the chemical potential difference between the bulk phases, $\gamma$ represents the surface tension, and $E_{strain}$ signifies the positive misfit strain free energy per unit volume, which depends on the elastic modulus of the solids. Finally, $E_{defect}$ refers to the energy of pre-existing defects per unit volume V.

\vfill

\section{\label{sec:level1} Discussion and Conclusion\protect\\ }
In conclusion, our study offers new insights into the atomistic mechanism for the solid-solid phase transitions in CsPbI\textsubscript{3} perovskites. Numerous X-ray crystallography experiments investigating the synthesis process have already established the active and crucial involvement of edge-sharing crystalline intermediate structures in the overall crystallization of perovskite thin-films. From simulations of direct phase transition from edge-sharing non-perovskite to perovskite structure, we find the formation of intermediate mixed edge-corner sharing structures that commonly identified as stacking faults or dislocation in wide range of perovskites. This knowledge is complemented by earlier findings from transmission electron microscopy (TEM) experiments\cite{chen_atomic_2022}, which have identified analogous intermediate edge-sharing stack-faulted domains in halide perovskites. Most importantly, photoemission electron microscopy based experimental investigations\cite{kosar_unraveling_2021} have further elucidated the presence of edge-dislocations and stack-faulted structures is harmful to the stability of halide perovskite under operational conditions. Recent experiments also revealed their possible photoplastic\cite{li_harnessing_2023, orr_imaging_2023, ducati_visualising_2023} nature of such structures, and signifying their capability to undergo movement under the influence of light during solar cell operations, therefore highlighting them as possible notorious degradation centers for PSCs. \\

From our results and analysis, one can conclude that the synthesis methodologies that involve the direct heating of the $\delta$-phase to perovskite can give rise to the formation of these defected intermediate structure and potentially stable on free energy surface. Hence, a careful attention is required in the development of large-scale industrial processes for producing PSCs those involving the direct annealing of the $\delta$-phase to perovskite. Secondly, our extensive investigation through large-scale seeded simulations, have unveiled crucial finding of the preferred occurrence of \textbf{nucleation} on a macroscopic level, characterized by a continuous transition marked by the formation of well-defined faceted perovskite crystals. Significantly, our simulations reveal the accelerated growth of the (100)-facet, which efficiently cut through the $\delta$-phase. \\

We observed that (100)-oriented seeds promoting the formation of the (100) facet could strategically optimize the perovskite crystallization process. Earlier experiments\cite{liu_high-performance_2017, lu_vapor-assisted_2020, lee_solid-phase_2020} on FAPbI\textsubscript{3} demonstrated that the controlled solid-solid phase transitions have produced highly efficient and stable PSCs. Building on our current study and observations of layer-by-layer transformation via a highly mobile interface, we propose two industrial scale synthesis methodologies to crystallize defect-free, and highly crystalline thin-films for PSCs and perovskite LEDs. First, strategies of surface vapor treatment-based can be devised to produce highly crystalline thin-film PSCs by a top to bottom layer-by-layer transformation as demonstrated by earlier experiments on FAPI\textsubscript{3}. Alternatively, molecular additives and chlorine can be employed, for instance, to initially create (100)-oriented domains of quasi-2D perovskite or Cs\textsubscript{2}PbI\textsubscript{2}Cl\textsubscript{2} serving as seeds for the uniform conversion of the $\delta$-phase to the complete perovskite structure. Therefore, simulations can ultimately contribute to the production of stable perovskite electronics. \\


Moreover, we provide valuable insights into how a potentially stable boundary may form between the (110)-facet of the perovskite crystal and its corresponding $\delta$-phase. Specifically, it may occur when the perovskite crystal reaches a critical size. Drawing inspiration from previous work by Haibo Zeng and co-workers\cite{chen_efficient_2021}, who engineered co-existing $\alpha$/$\delta$-CsPbI\textsubscript{3} thin-films to obtain all-inorganic white LEDs\cite{wani_advances_2023}, our study provides a platform to understand the impact of different process conditions for example effect of temperatures on free energy barriers for stabilizing the $\alpha$/$\delta$-interface. \\


\textbf{Therefore, we open avenues for enhancing the efficiency and stability of PSCs and perovskite LEDs building from all-atom MD simulations.}\\

Our study has been limited. Specifically, we have not fully explored the formation and significance of more complex structures, such as Cs\textsubscript{4}PbI\textsubscript{6}\cite{wang_formation_2021}, which could play a crucial role in the synthesis of perovskites. Therefore, looking ahead, our research road-map involves exploring the influence of non-stoichiometric complex structures, seed shapes and commonly used additives on phase transitions through precise finite temperature molecular dynamics studies of nucleation, utilizing large scale machine learning potentials derived from accurate quantum chemical calculations\cite{booth_fermion_2009, werner_efficient_2011, nagy_optimization_2018, gruber_applying_2018, cui_systematic_2022, bogdanov_enhancement_2022, ochi_tc++:_2023}. Accurate calculations of free energy barriers for the multi-step crystallization process will serve as a crucial parameter for calculating the rates of phase transitions to determine the actual kinetic stability of these meta-stable perovskites: one of the key unsolved problem for halide perovskites. \\

We have taken elementary steps in understanding the complex crystallization of multi-species materials and towards the design of atom by atom industrial process design, however importantly, the \textbf{first-principles-inspired all-atom computational framework} presented in this study offers promising opportunities for investigating complex structural transformations across a diverse range of multi-species materials. Examples include ferroelectric\cite{garcia_ferroelectric_2010, zhong_first-principles_1995, hwang_emergent_2012, manipatruni_scalable_2019, manipatruni_voltage_2018, mannhart_oxide_2010, zhang_ferroelectricity_2022, kimura_magnetic_2003}, multiferroics\cite{wang_epitaxial_2003}, and superconducting\cite{ohtomo_high-mobility_2004, reyren_superconducting_2007, bednorz_perovskite-type_1988, he_superconductivity_2001, bert_direct_2011, lee_linear--temperature_2023} materials with their potential to reduce global electricity consumption\cite{datta_electronic_1990, jones_how_2018, andrae_global_2015, ramesh_materials_2022}.



\vfill{ }

\section{\label{sec:level1}Methods\protect\\ }
\subsection{\label{sec:level1}MD simulations with RCPs\protect\\ }
\begin{figure}
  \includegraphics[width=0.80\linewidth]{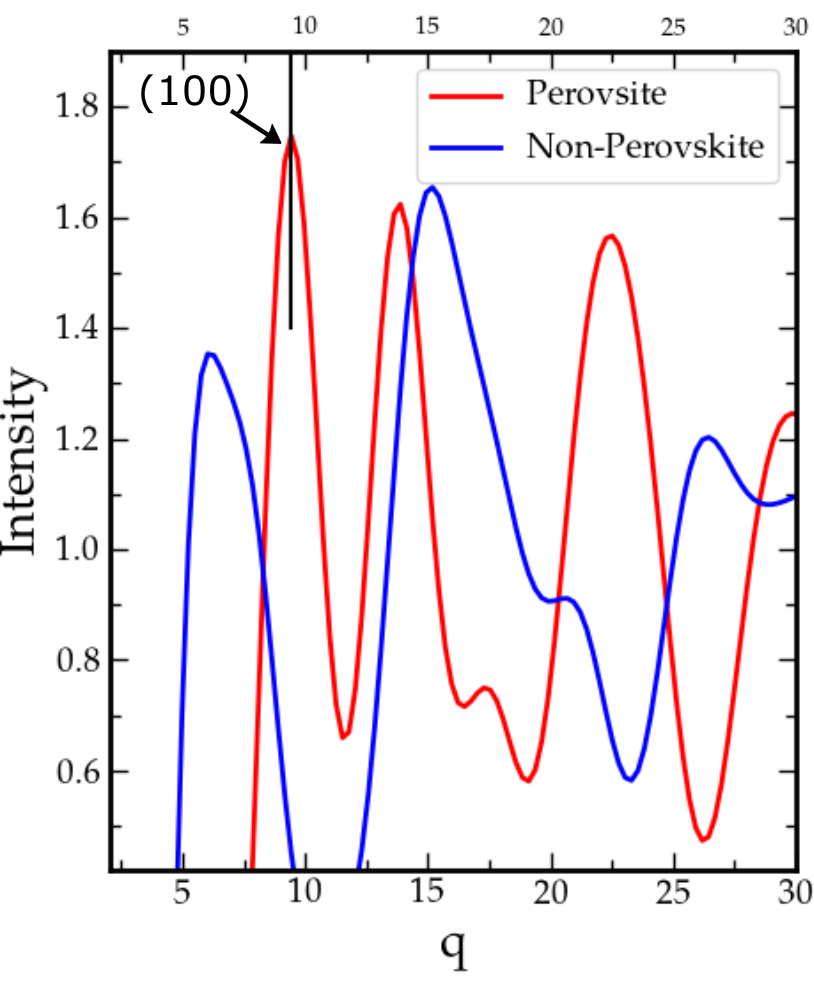}
  \caption{\label{fig:sf_m} Debye structure factors for perovskite($\alpha$) and non-perovskite($\delta$) phases of CsPbI\textsubscript{3}.}
\end{figure}

We utilized a reduced point charge\cite{zeron_force_2019, blazquez_madrid-2019_2022} inter-atomic potential for CsPbI\textsubscript{3} following the descriptions provided in the references\cite{bischak_liquid-like_2020}. The production runs are carried out in the isothermal-isobaric ensembles, using the Large-scale Atomic/Molecular Massively Parallel Simulator (LAMMPS) code (version 31 Mar 2017)\cite{thompson_lammps_2022}. Non-bonded interactions were restricted by a 1.0 nm cutoff, and electrostatic interactions were managed using the particle-particle-particle-mesh Ewald method. Temperature control was achieved using velocity re-scaling thermostat\cite{bussi_canonical_2007} with a relaxation time of 0.1 ps, while pressure control utilized a Parrinello-Rahman barostat\cite{martyna_constant_1994, parrinello_polymorphic_1981, shinoda_rapid_2004} with a relaxation time of 10 ps. OPES and umbrella sampling biased simulations were conducted using PLUMED 2.8\cite{tribello_plumed_2014, bonomi_plumed:_2009}, interfaced with LAMMPS. To scrutinize finite-size effects concerning yellow to black phase transitions, we employed supercells of varying sizes, ranging from 32, 128, 512 to 1024 formula units (f.u.) of $\delta$-CsPbI\textsubscript{3}. In the context of seeded simulations, we started by generating three distinctive configurations featuring spherical perovskite seed sizes with radii of 3.7 nm, 5.0 nm, and 7.0 nm. These seeds were embedded within $\delta$-phase structures, characterized by approximate dimensions of 16x16x16 nm (88540 atoms), 23x23x23 nm (272310 atoms), and 33x31x33 nm (675180 atoms), respectively. It is crucial to emphasize that the selection of the simulation box size, representing $\delta$-phases, was purposefully set to be more than at least four times the radius of the perovskite seeds. This consideration ensures that the perovskite seeds do not interact with each other within the confines of periodic boundary conditions in MD simulations. Additionally, meticulous attention was given to maintaining suitable distances between pairs of atoms, aligning with the Pb–Pb, Pb-Cs, and Pb-I distances between the two phases. Subsequent to the observation of perovskite seed growth and dissolution in these three initial systems, we expanded our configurations. Additional simulations were conducted, introducing perovskite seeds with radii of 4.0 nm, 4.5 nm, 4.8 nm, 5.0 nm, 5.3 nm, 5.5 nm, and 6.0 nm into a simulation box with dimensions approximately 33x31x33 nm, encompassing around total 678770 atoms. These subsequent simulations were intentionally designed first to repeat earlier simulations, and to offer qualitative insights into the associated free energy barrier for the process. The structure factors were computed employing a methodology similar to previous studies\cite{ahlawat_combined_2021}, and are commonly depicted, as shown in Figure \ref{fig:sf_m}, for different polymorphs. To facilitate analysis of the crystallization process, we have normalized the intensities relative to the initial configurational intensities, thus initiating values from 1.0 in Figures \ref{fig:seed_sf} and \ref{fig:seed_sfg}.

\subsubsection{\label{sec:level1}Melting point\protect\\ }
\begin{figure}
  \includegraphics[width=\linewidth]{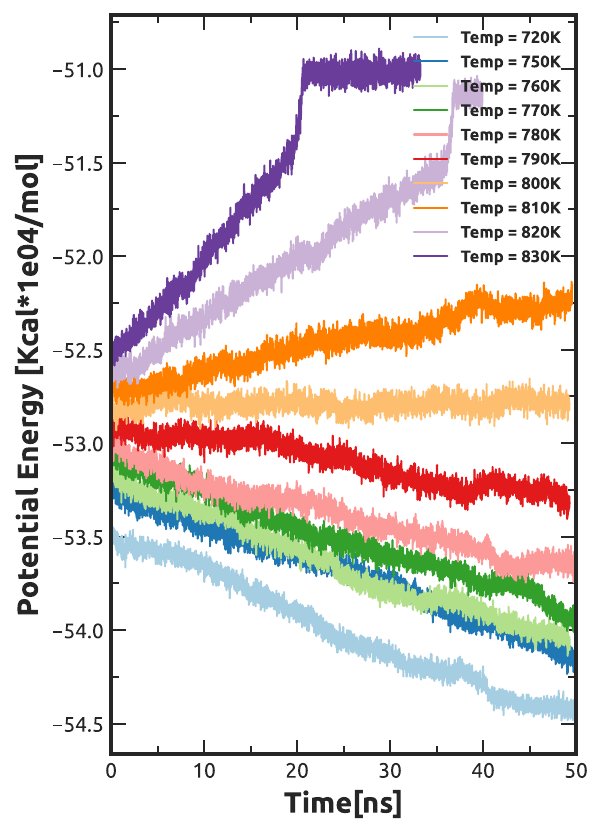}
  \caption{\label{fig:melting_p} The figure shows the time evolution of potential energy in co-existing simulations at different temperatures.}
\end{figure}

To determine the melting point of CsPbI\textsubscript{3} using the reduced charge inter-atomic potential, we carried out co-existing simulations\cite{ten_wolde_simulation_1996, garcia_fernandez_melting_2006} involving crystalline seeds composed of 6400 atoms in contact with a melt containing equal number of atoms. Subsequently, isothermal-isobaric simulations were executed over a temperature range spanning from 720 to 850 K. The progression of potential energy at these temperatures is illustrated in Figure \ref{fig:melting_p}, indicating that perovskite melts above 800 K and grows below 800 K. It is noteworthy that the experimental melting point of CsPbI\textsubscript{3} is around 750 K\cite{straus_kinetically_2019}.

\subsubsection{\label{sec:level1}local q4 analysis\protect\\ }
\begin{figure}
  \includegraphics[width=\linewidth]{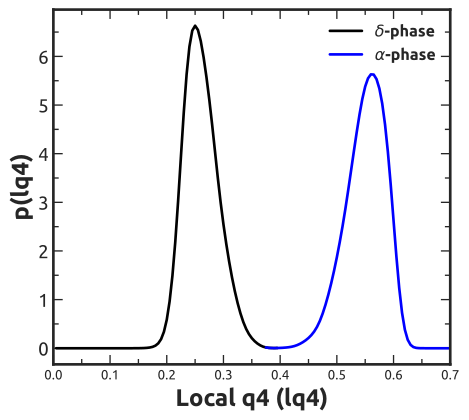}
  \caption{\label{fig:lq4_histo} Histogram illustrating the distribution of $q_4$ for a combined simulated trajectory of $\alpha$ and $\delta$ phases.}
\end{figure}

\begin{figure}
  \includegraphics[width=\linewidth]{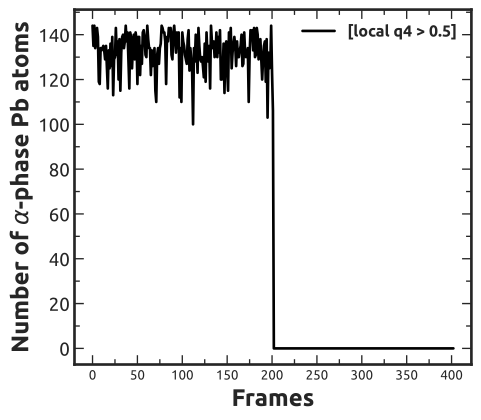}
  \caption{\label{fig:lq4_cd}Time evolution of local $q_4$ exceeding a threshold of 0.5 in a combined simulated trajectory featuring both $\alpha$ and $\delta$ phases.}
\end{figure}

To quantify the number of formula units within the perovskite phase, whether in direct coexistence simulations or clusters in seeding simulations, we employ the rotationally invariant local-bond order parameter, denoted as lq4\cite{rein_ten_wolde_numerical_1996, van_duijneveldt_computer_1992, lechner_accurate_2008, tribello_plumed_2014, bonomi_plumed:_2009}. This parameter is determined by the relative positions of a tagged particle and its neighbors within a defined distance. In our investigation, we restrict the calculation of lq4 to the Pb species within the systems, considering a cutoff distance of 8 \AA. Figure \ref{fig:lq4_histo} illustrates the histogram of lq4 values of a finite-temperature simulated trajectories for perovskite and non-perovskite phases. Notably, a threshold can be established at lq4 $>$ 0.5, enabling the calculation of the number of perovskite formula units in a given system. To further validate this approach, we conduct a simple test on a trajectory created by combining an equal number of frames from simulated perovskite and non-perovskite phases. As depicted in Figure \ref{fig:lq4_cd}, lq4 effectively distinguishes and counts only the number of perovskite units. Similarly, the quantity of $\delta$-phase can be assessed by considering lq4 $<$ 0.33. 


\subsection{\label{sec:level1}RPA and DFT calculations\protect\\ }
In this study, all DFT and RPA calculations were executed using the VASP code\cite{kresse_efficiency_1996, kresse_efficient_1996}. The projected Aagmented wave functions (PAW) with PBE potentials\cite{kresse_ultrasoft_1999} were employed for DFT calculations, and LDA potentials were used for LDA functional. RPA+HF results were also obtained using PBE PAW potentials, given their comparability with those derived from PAW GW potentials\cite{braeckevelt_accurately_2022}. We systematically compared various exchange-correlation (XC) functionals and dispersion methods, including LDA\cite{hohenberg_inhomogeneous_1964, kohn_self-consistent_1965}, PBE\cite{perdew_generalized_1996}, PBEsol\cite{csonka_assessing_2009}, r2SCAN\cite{furness_accurate_2020}, PBE+D3(BJ)\cite{grimme_consistent_2010, grimme_effect_2011, johnson_post-hartree-fock_2006, becke_density-functional_2005},  r2SCAN+rVV10\cite{ning_workhorse_2022}, optB86b\cite{klimes_van_2011} and RPA+HF\cite{harl_assessing_2010, bohm_collective_1951, gell-mann_correlation_1957, kaltak_cubic_2014, braeckevelt_accurately_2022}. Each polymorph was represented using 20-atom supercells, and a cutoff energy of 600 eV was established. Different k-point grids of 4x10x2, 6x6x4, 6x6x4, 4x6x4, and 6x6x4 for DFT calculations and for RPA+HF calculations k-point grids of 2x6x1, 2x3x2, 2x3x2, 2x3x2, 2x3x3 were used for $\delta$, $\gamma$, $\beta$, $\alpha$, and manually constructed face-sharing phase, respectively. AIMD\cite{payne_iterative_1992} simulations using k-points were conducted to generate the training dataset, utilizing the r2SCAN+rVV10 functional. These simulations involved 40-atom supercells for each polymorph and were carried out within the isothermal-isobaric (NPT) ensemble, employing a 2fs time step. Pressure and temperatures were controlled using the Parrinello-Rahman barostat and Langevin thermostat\cite{10.1093/oso/9780198803195.001.0001}, respectively. \\



\subsection{\label{sec:level1} Machine Learning Potentials\protect\\ }

\subsubsection{\label{sec:level1}Nequip MLPs\protect\\ }
The Nequip\cite{batzner_e3-equivariant_2022} MLPs are derived from AIMD simulations data based on r2SCAN+rVV10 DFT level of theory, as described in above sections. For the training process, we employed three interaction layers, implementing a radial cutoff of 8 \AA and embedding interatomic distances in a basis comprising eight Bessel functions. Feature representations were selected up to rotation orders of l=2. The loss function incorporated forces, energies, and stresses obtained from DFT calculations, and optimization was executed using Adam with a learning rate of 0.003 and a batch size of 10. The training procedure was conducted on a V100 GPUs.

\subsubsection{\label{sec:level1}NEP-GPUMD simulations\protect\\ }
For large-scale simulations using MLPs, we utilized the neuroevolution potentials (NEPs)\cite{fan_neuroevolution_2021, fan_improving_2022} model in GPUMD code\cite{fan_gpumd:_2022, fan_efficient_2017}, as extensively discussed recent publications\cite{fransson_phase_2023, fransson_revealing_2023}. NEPs feature a neural network architecture where local atomic environments are delineated by radial and angular components. The radial component of the atomic environment descriptor is designed through linear combinations of Chebyshev basis functions, while the three-body angular part is similarly constructed from Legendre polynomials. In this study, we tried two different setups, first with a radial and angular cutoffs at 6\AA and 5\AA, respectively, and second with using radial and angular cutoffs cutoffs at 8\AA and 4\AA, respectively, both with radial and angular expansions at orders 12 and 8. The neural network architecture consisted of hidden layer comprising 50 neurons, and the training process involved 100,000 generations. The ultimate model for CsPbI\textsubscript{3} was trained based on forces, energies, and stresses derived from r\textsuperscript{2}SCAN+rVV10 DFT calculations. In order to perform co-existing solid-solid simulations, we prepare configurations comprising different polymorphs with a minimal lattice mismatch between them. Specifically, for the case of (100)$_{\delta}\|(100)_{\alpha}$ , a supercell consisting of 2520 atoms (with dimensions of 88.046 x 37.734 x 37.734 \AA) representing the $\alpha$-phase is combined with a 2560-atom supercell (with dimensions of 83.472 x 38.320 x 35.522 \AA) representing the $\delta$-phase. The final configuration is attained by averaging the lattice parameters, resulting in dimensions of 175.0 x 37.847 x 36.448 \AA. For the case of (100)$_{\delta}\|(110)_{\alpha}$ , we first create the (110)-facet of the $\alpha$-phase utilizing ase utilities. Subsequently, it is combined with a supercell of the $\delta$-phase, ensuring minimal mismatch between the two structures. All structures are provided in the supplementary materials.


\newpage

\section*{Acknowledgements}
This research is funded by Swiss National Science Foundation through Post.Doc mobility fellowship $P500PN\_206693$. I am sincerely thankful to Professors Daan Frenkel, Michael Graetzel, Nitin Padture, Ali Alavi, Volker Heine, and Carlos Vega for helpful discussions, encouragements, and useful guidance. I thank Professors Erin Johnson, Michael Sprik, Gabor Csyani, and Alessandro Laio for helpful insightful discussions. I am greatly thankful to Dr. Felix Musil for discussion and help regarding SOAP kernels. I thank Dr. Michele Invernizzi for discussions on enhanced sampling simulations. I am greatly thankful to Dr. Jiahuan Zhang, Dr. Zaiwei Wang, Dr. Chen Li and Dr. Pepe Marquez for highly insightful discussions on experimental data of phase transitions between $\delta$-$\alpha$ phase transitions. I thank Ben Shi and Dr. Michele Simoncelli for carefully reading the manuscript and providing helpful comments. I am indebted and greatly thankful to Yusuf Hamied Department of Chemistry, Professor Angelos Michaelides for hosting and providing necessary computational resources through Rogue-GPU cluster, Nest-CPU cluster, UK Materials and Molecular Modelling Hub YOUNG CPU/GPU cluster, which is partially funded by EPSRC (EP/T022213/1, EP/W032260/1 and EP/P020194/1) and ARCHER2 UK National Supercomputing Service.
\section*{Data availability}
All parts of this study are easily reproducible: all kinds of input/output files, starting configurations, notified output configurations, plumed input files, lammps input files, RPA/DFT input files and MD trajectories are available on open access server Zenodo: \url{10.5281/zenodo.10938457}, Github: \url{https://github.com/ahlawat-paramvir/plumed2}, \url{https://github.com/ahlawat-paramvir/CsPbI3_phase_transitions} and also from author.

\nocite{*}

\onecolumngrid

\section*{References}

\twocolumngrid

\bibliography{apssamp}


\end{document}